\documentclass[11pt, a4paper]{article}
\usepackage{jheparxiv}
\usepackage[latin1]{inputenc}
\usepackage{amsmath}
\usepackage{amsfonts}
\usepackage{amssymb}
\usepackage{latexsym}
\usepackage{mathrsfs}
\usepackage{graphicx}
\usepackage{color}
\usepackage{slashed}
\usepackage{twistor}
\usepackage[all]{xy}

\renewcommand{\d}{\mathrm{d}}

\newcommand{\sM}{\mathsf{M}}
\newcommand{\m}{\mathfrak{m}}
\newcommand{\Zt}{\mathcal{Z}}

\subheader{
\begin{flushright}
\texttt{IMPERIAL-TP-TA-2016-01}\\
 \hfill \texttt{TCDMATH 16-16}
\end{flushright}}

\title{Conformal higher spin scattering amplitudes from twistor space}

\author[a]{Tim Adamo,}
\author[b]{Philipp H\"ahnel}
\author[b]{\& Tristan McLoughlin}

\affiliation[a]{Blackett Laboratory \\
        Imperial College \\
        London, SW7 2AZ, United Kingdom}

\affiliation[b]{School of Mathematics \\
        Trinity College Dublin \\
        College Green, Dublin 2, Ireland}

\emailAdd{t.adamo@imperial.ac.uk}
\emailAdd{[haehnel,tristan]@maths.tcd.ie}
\abstract{We use the formulation of conformal higher spin (CHS) theories in twistor space to study their tree-level scattering amplitudes, finding expressions for all three-point $\overline{\mbox{MHV}}$ amplitudes and all MHV amplitudes involving positive helicity conformal gravity particles and two negative helicity higher spins. This provides the on-shell analogue for the covariant coupling of CHS fields to a conformal gravity background. We discuss the restriction of the theory to a ghost-free unitary subsector, analogous to restricting conformal gravity to general relativity with a cosmological constant. We study the flat-space limit and show that the restricted amplitudes vanish, supporting the conjecture that in the unitary sector the S-matrix of CHS theories is trivial. However, by appropriately rescaling the amplitudes we find non-vanishing results which we compare with chiral flat-space higher spin theories.}

\begin{document}
 
\maketitle

\section{Introduction}

The existence of interacting field theories with higher spins in $d>3$ is tightly constrained by a variety of no-go theorems. Indeed, it seems that there are only two such higher-spin theories which are generally agreed to be well-defined: Vasiliev's theory in space-times with a non-vanishing cosmological constant~\cite{Vasiliev:1990en,Vasiliev:2003ev}, and \emph{conformal higher spin} (CHS) theories~\cite{Fradkin:1985am}. CHS theories can be thought of as higher-spin generalizations of conformal gravity and as such, they are non-unitary theories whose equations of motion involve higher derivatives of the underlying gauge fields. Despite this obvious lack of unitarity, there are several reasons why these theories are interesting.

Over the last thirty years, CHS theory has been extended to an interacting theory involving single copies of conformal fields at \emph{all} integer spins $s\geq1$~\cite{Fradkin:1989md,Segal:2002gd}. Both conformal gravity and CHS theories play an interesting role in the study of other CFTs. Conformal gravity appears in the induced action upon coupling $\cN=4$ super-Yang-Mills to background conformal supergravity and integrating out the SYM fields~\cite{Liu:1998bu,Tseytlin:2002gz}. CHS theory similarly appears in the UV divergent part of the effective action found by minimally coupling the free O$(N)$ vector model in $d=4$ to an infinite set of higher spin symmetry currents~\cite{Tseytlin:2002gz,Segal:2002gd,Bekaert:2010ky}; in fact this is often the most practical way of computing the non-linear CHS interactions. %The free O$(N)$ vector model in 4-dimensions is itself the conjectured holographic dual~\cite{Klebanov:2002ja} to Vasiliev's higher spin theory in AdS$_5$~\cite{Vasiliev:2003ev}. 

The infinite-dimensional higher-spin conformal symmetry should constrain CHS theory to be renormalizable and even UV-finite at the quantum level. In fact it has been shown that the free, one-loop partition function on a Minkowski~\cite{Beccaria:2015vaa} and 4-sphere~\cite{Giombi:2013yva,Tseytlin:2013jya} backgrounds vanishes, which indicates that, for a suitably chosen regularisation, there is a remarkable cancellation between physical degrees of freedom. Despite these intriguing features and the various non-perturbative calculations that have been done, relatively little is known about the perturbative observables such as scattering amplitudes in CHS theory. At a practical level, this is due to the difficulty of determining interacting terms in the Lagrangian, defined as an induced theory. 

Nevertheless, recent progress has been made in calculating the four-point tree-level scattering amplitudes of CHS theory with external scalars~\cite{Joung:2015eny}, gluons or gravitons~\cite{Beccaria:2016syk}. In these calculations, external states are chosen to be solutions of the linearized two derivative -- or \emph{unitary} -- equations of motion, which form a consistent subset of the linearized CHS equations of motion. When all CHS intermediary states are summed over, the resulting four-point amplitudes vanish. It has been conjectured that this vanishing should extend to any number of unitary external states, order-by-order in perturbation theory: in other words, the S-matrix of CHS theory (defined in this way) is trivial.

Given the complexity of the space-time action, it seems a difficult task to prove this conjecture even at tree-level, however. An alternative approach is offered by \emph{twistor theory}, which is a natural formalism for studying any four-dimensional theory with conformal symmetry.\footnote{Recall that a twistor space is a three-dimensional complex manifold $\CPT$ related non-locally to a self-dual, four-dimensional space-time $M$: each point $x\in M$ corresponds to a holomorphically embedded Riemann sphere $X\cong\CP^{1}\subset \CPT$.} In~\cite{Haehnel:2016mlb}, two of us showed that linear CHS theory can be described by action functionals in twistor space and gave a conjectural twistor action for the full interacting theory
 \footnote{There are many interesting connections between twistors and higher spin theory. For example, in \cite{Gelfond:2008td}  a twistor-like interpretation of the $Sp(8)$ invariant formulation of massless fields in four-dimensions was given. It would be very interesting to better understand the relation between this formulation and that discussed in this work. We thank M. Vasiliev for bringing this to our attention.}. The object of this paper is to apply that twistor action to the computation of some tree-level scattering amplitudes in CHS theory.

\medskip

After reviewing the fact (perhaps not widely known) that CHS theory admits a perturbative expansion around a self-dual sector, the twistor action construction is outlined in section~\ref{TAR}. We also define the unitary subsector in twistor space by introducing an additional structure (the infinity twistor) which breaks conformal invariance. Section~\ref{CHSamp} computes two (infinite) families of tree-level scattering amplitudes with the twistor action. These are three-point amplitudes with two positive and one negative helicity external states of arbitrary spin ($\overline{\mbox{MHV}}$ amplitudes), and $n$-point amplitudes with two spin $s$ negative helicity states and $(n-2)$ positive helicity states of spin two (MHV amplitudes on a conformal gravity background). In the first instance these amplitudes are defined with arbitrary linearized external states. Section~\ref{Uamp} gives the amplitudes with unitary external states; these are seen to vanish in Minkowski space, supporting the conjecture of~\cite{Joung:2015eny,Beccaria:2016syk}. If the amplitudes are re-scaled by spin-dependent powers of the cosmological constant before taking the flat space limit, then interesting non-vanishing answers are obtained. Considering the unitary restriction of the self-dual sector in twistor space suggests the existence of certain chiral, non-conformal higher spin theories. We compare these results with other attempts to construct flat space higher spin theories in the discussion of section~\ref{Dis}.

%%%%%%%%%%%%%%%%%%%%%%%%%%%%%%%%%%%%%%%%%%%%%%%%%%%%%%%%
%%%%%%%%%%%%%%%%%%%%%%%%%%%%%%%%%%%%%%%%%%%%%%%%%%%%%%%%

\section{Conformal Higher Spins on Twistor Space}
\label{TAR}

Conformal higher spin (CHS) theories were first formulated at the linearized level in terms of rank $s$ symmetric fields $\phi_{\mu_1\cdots\mu_s}$ with free action~\cite{Fradkin:1985am}
\be\label{linCHS1}
S^{s}[\phi]=\frac{1}{\varepsilon^2}\int \d^{4}x\,\phi_{\mu(s)}\,P^{\mu(s)\,\nu(s)}\,\phi_{\nu(s)}\,,
\ee
where $\varepsilon$ is a dimensionless coupling parameter and we use the multi-index notation $\phi_{\mu(s)}\equiv\phi_{(\mu_{1}\cdots\mu_{s})}$ for totally symmetric indices. The operator $P^{\mu(s) \,\nu(s)}$ is of order $2s$ in space-time derivatives and obeys $P^{\mu(s)\,\nu(s)}=P^{\nu(s)\,\mu(s)}$, $P_{\mu}^{\;\mu\mu(s-2)\,\nu(s)}=0$, and $P^{\mu\mu(s-1)\,\nu(s)}\partial_{\mu}=0$. A key motivation for introducing the  operator $P^{\mu(s) \,\nu(s)}$ is that it provides a projection onto pure spin-$s$ states and the resulting action possess a maximal degree of gauge invariance and irreducibility even off the mass shell. The action \eqref{linCHS1} is invariant under both the local (differential) and algebraic (Weyl) gauge symmetries
\begin{equation*}
 \delta\phi_{\mu_1\cdots\mu_s}(x) = \partial_{(\mu_1}\epsilon_{\mu_2\cdots\mu_s)}(x)- \delta_{(\mu_1\mu_2} \alpha_{\mu_3\cdots\mu_s)}(x)
 % -\mathrm{traces}
\end{equation*}
for totally symmetric  $\epsilon_{\mu(s-1)}(x)$ and $\alpha_{\mu(s-2)}(x)$. These free theories were extended to cubic interactions some time ago~\cite{Fradkin:1989md,Fradkin:1990ps}, and can be completed to an interacting CHS theory involving single copies of conformal fields at \emph{all} spins for integer $s\geq0$~\cite{Tseytlin:2002gz,Segal:2002gd,Bekaert:2010ky}.

\medskip

In this section, we review the formulation of CHS theory in \emph{twistor space}~\cite{Haehnel:2016mlb}, making use of a perturbative expansion around a self-dual (SD) sector. We also describe a unitary subsector of CHS theory, and discuss the various ways that external states of arbitrary spin are represented in twistor space.

%%%%%%%%%%%%%%%%%%%%%%%%%%%%%%%%%%%%%%%%%%%%%%%%%%%%%%%

\subsection{Perturbative expansion around a self-dual sector}

For any fixed spin $s$, the free CHS action \eqref{linCHS1} can be written in terms of a (linear) generalized Weyl `curvature' tensor, $C^{(s)}_{\mu(s) \nu(s)}$ which is totally symmetric and trace-free in each $s$-tuple of indices, and antisymmetric between the two~\cite{Fradkin:1989md,Fradkin:1990ps}, see also \cite{Marnelius:2008er,Vasiliev:2009ck}. Schematically, the relation between the Weyl curvatures and the higher-spin gauge fields is $C^{(s)}\sim \cP \partial^{s}\phi$, where $\cP$ is a projector enforcing the appropriate symmetries. The free CHS action is the natural generalization of the $s=1$ Maxwell action for electromagnetism or the $s=2$ Weyl action for conformal gravity:
\be\label{linCHS2}
S^{s}[\phi]=\frac{(-1)^s}{2\varepsilon^2}\int \d^{4}x\,C^{(s)}_{\mu(s)\,\nu(s)}\,C^{(s)\,\mu(s)\,\nu(s)}\,.
\ee
The interpretation of $C^{(s)}$ as a generalized Weyl curvature is meaningful, since $C^{(s)}=0$ if and only if the gauge field $\phi$ is conformally trivial~\cite{Damour:1987vm}. The equations of motion in these variables are a spin $s$ extension of the Bach equations for conformal gravity.%: $B^{(s)}_{\mu(s)}\sim \nabla^{\nu(s)} C^{(s)}_{\mu(s)\,\nu(s)}$.

In $d=4$, standard conformal curvature tensors (such as $F_{\mu\nu}$ for a gauge field or $C^{(2)}_{\mu\nu\rho\sigma}$ for a metric) admit a decomposition into self-dual (SD) and anti-self-dual (ASD) parts, since the Hodge star acts on 2-forms as an involution. The same is true for the generalized Weyl curvature tensors; equating a 4-vector index with a pair of $\SL(2,\C)$ Weyl spinor indices $A,A'$, this decomposition reads:
\be\label{SDd}
C^{s}_{\mu(s)\,\nu(s)}=C^{(s)}_{A(s) A'(s)\,B(s) B'(s)}=\epsilon_{A(s) B(s)}\, \widetilde{\Psi}_{A'(s) B'(s)} + \epsilon_{A'(s) B'(s)}\,\Psi_{A(s)B(s)}\,,
\ee
where $\epsilon_{A(s) B(s)}:= \epsilon_{A_1 B_1}\epsilon_{A_2 B_2}\cdots \epsilon_{A_s B_s}$. The spinors $\widetilde{\Psi}_{A'(s) B'(s)}$, $\Psi_{A(s)B(s)}$ are totally symmetric and encode the SD and ASD curvature of the spin $s$ gauge field, respectively. The equality between $C^{s}$, whose representation of SO$(4)$ corresponds to a two-row Young tableau with $2(2s+1)$ components, and the totally symmetric representations $\Psi$ and $\tilde \Psi$, each with $2s+1$ components, is a special feature of four-dimensions. The linearized equations of motion for these spinors are an extension of the Bach equations:
\be\label{chseom1}
\nabla^{A(s) A'(s)}\Psi_{A(s) B(s)}= 0 =\nabla^{A(s) A'(s)} \widetilde{\Psi}_{A'(s) B'(s)}\,,
\ee
with equivalence between the two conditions following from the Bianchi identities for $C^{(s)}$.

%Extending the free theory to the interacting one entails replacing ordinary space-time derivatives by covariant derivatives with respect to the full tower of CHS gauge fields, for $s=0,1,2,\ldots$. At each spin, the decomposition \eqref{SDd} should still hold, but with the spinors $\widetilde{\Psi}_{A'_S B'_S}$, $\Psi_{A_S B_S}$ now non-linear functions of \emph{all} the $\phi$s -- not just the spin $s$ one. Similarly, the equations of motion \eqref{chseom1} will remain functionally the same on a conformally flat background after taking into account the non-linear connection terms contributing to the covariant derivatives.

From the functional form of the equations of motion \eqref{chseom1}, we can immediately identify two interesting subsectors of solutions to the free CHS equations of motion. The first is the \emph{self-dual (or instanton) sector}\footnote{Of course, the ASD sector is also a valid solution to CHS theory.}, defined by
\be\label{SDsector}
\Psi_{A(s) B(s)}=0\,.
\ee
The second is the \emph{unitary subsector}, defined by the zero-rest-mass equations for helicity $+s$ and $-s$:
\be\label{zrm1}
\nabla^{AA'}\widetilde{\Psi}_{A' C'(s-1)B'(s)}=0\,, \qquad \nabla^{AA'}\Psi_{AC(s-1)B(s)}=0\,.
\ee
These are second-order equations for the higher-spin gauge fields, which are equivalent to the Fronsdal equations for massless higher spin fields.

\medskip

Consistency of the SD sector can be used to provide an interesting perturbative expansion for CHS theories, analogous to the Chalmers-Siegel action for Yang-Mills theory~\cite{Chalmers:1996rq,Chalmers:1997sg} or the Berkovits-Witten action for conformal gravity~\cite{Berkovits:2004jj}. In terms of the SD and ASD spinors, the linearized action \eqref{linCHS2} is
\be\label{linCHS3}
S^{s}[\phi]=\frac{1}{2\varepsilon^2}\int \d^{4}x\,\left(\Psi_{A(s) B(s)} \Psi^{A(s) B(s)} +\widetilde{\Psi}_{A'(s) B'(s)} \widetilde{\Psi}^{A'(s) B'(s)}\right)\,,
\ee
up to an overall numerical factor. In turn, this is trivially equal to
\be\label{linCHS4}
\frac{1}{\varepsilon^2}\int \d^{4}x\,\Psi_{A(s) B(s)} \Psi^{A(s) B(s)} + \frac{1}{2\varepsilon^2}\int \d^{4}x\,\left(\widetilde{\Psi}_{A'(s) B'(s)} \widetilde{\Psi}^{A'(s) B'(s)} - \Psi_{A(s) B(s)} \Psi^{A(s) B(s)}\right)\,.
\ee
The combination $(\widetilde{\Psi}^2-\Psi^2)$ is a total derivative 
\footnote{One can write the field strengths in terms of a potential
  \be
 \Psi_{A(s)B(s)}=\nabla^{A'_1}{}_{(A_1} \dots \nabla^{A'_s}{}_{A_s}\phi_{B(s))A'(s)}~,~~~{\rm and}~~~
\tilde{\Psi}_{A'(s)B'(s)}=\nabla^{A_1}{}_{(A'_1}\dots \nabla^{A_s}{}_{(A'_s}\phi_{B'(s))A(s)}~.
 \ee
Using the gauge condition $\nabla^{A_1A'_1}\phi_{A_1A(s-1)A'_1A'(s-1)}=0$ we can 
write this as 
 \be
 \Psi_{A(s)B(s)}=\nabla^{A'_1}{}_{A_1} \dots \nabla^{A'_s}{}_{A_s}\phi_{B(s)A'(s)}
 \ee
 and similarly for $\tilde{ \Psi}_{A'(s)B'(s)}$. 
For the linearised theory, where the derivatives are treated as commuting, one can compute that
   \begin{align}
 \Psi^{A(s)B(s)}\Psi_{A(s)B(s)}&= (\half)^s 
 \phi^{B(s)B'(s)}  (\nabla^2)^s \phi_{B(s)B'(s)}
 \end{align}
up to total derivative terms. Repeating the calculation for $\tilde{ \Psi}^{A'(s)B'(s)}\tilde{\Psi}_{A'(s)B'(s)}$
leads to the same result and so $
\Psi^{A(s)B(s)}\Psi_{A(s)B(s)}-\tilde{ \Psi}^{A'(s)B'(s)}\tilde{\Psi}_{A'(s)B'(s)}= 0$
 up to total derivative terms.}, so does not contribute to perturbative calculations and can be discarded. 

By introducing a purely ASD Lagrange multiplier $G_{A(s) B(s)}$, totally symmetric on all of its spinor indices, we arrive at the alternative linearized CHS action:
\be\label{linCHS5}
S^{s}[\phi, G]= \int \d^{4}x\,G_{A(s) B(s)}\,\Psi^{A(s) B(s)} - \frac{\varepsilon^{2}}{2} \int \d^{4}x\,G_{A(s) B(s)}\,G^{A(s) B(s)}\,.
\ee
The utility of this re-writing is obvious upon inspecting the equations of motion:
\be\label{chseom2}
\Psi_{A(s) B(s)}= \varepsilon^2\,G_{A(s) B(s)}\,, \qquad \nabla^{A(s) A'(s)}\Psi_{A(s) B(s)}= 0\,.
\ee
The coupling $\varepsilon$ now manifestly controls the expansion of the theory around the SD sector: when $\varepsilon=0$, $\Psi_{A(s) B(s)}=0$. The Lagrange multiplier $G_{A(s) B(s)}$ is interpreted as an ASD linearized spin-$s$ field propagating on this SD background.

%While the action \eqref{linCHS5} is defined for the linear CHS theory, we conjecture that this general structure (\textit{i.e.}, the existence of a perturbative expansion around the SD sector) exists for the full interacting theory as well.

%%%%%%%%%%%%%%%%%%%%%%%%%%%%%%%%%%%%%%%%%%%%%%%%%%%%%%%

\subsection{Twistor actions for CHS theory}

For a general SD space-time, $M$, the associated twistor space $\CPT$ is a complex 3-manifold obtained from a complex deformation of an open subset of $\CP^3$. The twistor space and its space-time are related non-locally: a point $x\in M$ corresponds to a holomorphically embedded Riemann sphere $X\cong\CP^1$ in $\CPT$, with the conformal structure of $M$ determining the complex structure on $\CPT$~\cite{Penrose:1976js,Atiyah:1978wi}. There are a variety of reviews on twistor theory (\textit{c.f.}, \cite{Penrose:1986ca,Ward:1990,Adamo:2011pv}); we mainly follow the notation of~\cite{Adamo:2013cra}. 

Let $(x^{AA'},\sigma_{B})$ be coordinates on the projective un-primed spinor bundle of $M$: $\PS^+\cong M\times\CP^1$. The non-local relationship between $M$ and $\CPT$ is captured by the \emph{incidence relations}
\be\label{incidence}
\Zt^{\alpha}:M\times\CP^{1}\rightarrow\CPT, \qquad \Zt^{\alpha}(x,\sigma)=\left(\mu^{A'}(x,\sigma)\,, \lambda_{A}(x,\sigma)\right)\,,
\ee
where the components of the map $\Zt^{\alpha}$ are (holomorphic) homogeneous coordinates, considered only up to scale. The map $\Zt^{\alpha}(x,\sigma)$ is homogeneous of degree one with respect to the coordinates $\sigma_A$ on $\CP^1$.

To describe a linear theory, it suffices to consider the `flat' twistor space $\PT$ of Min\-kows\-ki space-time, an open subset of $\CP^3$. In this case, points $x\in\M$ correspond to linearly embedded Riemann spheres and the incidence relations \eqref{incidence} are simply:
\begin{equation*}
 \mu^{A'}=\im x^{AA'}\lambda_{A}\,, \qquad \lambda_{A}=\sigma_{A}\,.
\end{equation*}
In the following we will denote coordinates on $\CPT$ by $\Zt^\alpha$ and coordinates on $\PT$ by $Z^\alpha$. The linear CHS action \eqref{linCHS5} was lifted to twistor space in~\cite{Haehnel:2016mlb}, resulting in a functional of two twistor fields:
%\be
%	f^{(s)} \in \Omega^{0,1}(\PT, \odot^{s} T^{1,0}(\PT))\,,
%	\qquad
%	g^{(s)} \in \Omega^{0,1}(\PT, \odot^{s}T^{* 1,0}(\PT))
%\ee
\be\label{tf}
%	f^{\alpha(s-1)}\in\Omega^{0,1}(\PT,\odot^{s-1} T_{\PT}(s-1))\,, \qquad g_{\alpha(s-1)}\in\Omega^{0,1}(\PT, \odot^{s-1}T^{*}_{\PT}(-s-3))\,,
	f^{\alpha(s-1)}\in\Omega^{0,1}(\PT,\cO(s-1))\,, \qquad g_{\alpha(s-1)}\in\Omega^{0,1}(\PT, \cO(-s-3))\,,
\ee
%using the notation $\cV(k)\cong\cV\otimes\cO(k)$ for some vector bundle $\cV$ and 
$\cO(k)$ being the sheaf of holomorphic functions homogeneous of degree $k$. This data is considered only up to the gauge transformations and constraints
\be\label{gt}
f^{\alpha_1\cdots\alpha_{s-1}}\rightarrow f^{\alpha_1\cdots\alpha_{s-1}} + Z^{(\alpha_1}\Lambda^{\alpha_2\cdots\alpha_{s-1})}\,, \quad Z^{\alpha_1}g_{\alpha_1\cdots\alpha_{s-1}}=0\,.
\ee 
For objects on twistor space we will from now on use the multi-index notation
\be
	f^{\alpha_I}\equiv f^{\alpha(|I|)}
\ee
also used in ~\cite{Haehnel:2016mlb}. The linearized CHS twistor action is: 
\be\label{linTCHS}
	S^{s}[f,g]
		= \int_{\PT} \!\!\D^{3} Z\, g_{\alpha_I}\wedge \dbar f^{\alpha_I} 
			- \frac{\varepsilon^{2}}{2}\int\limits_{\PT\times_{\M}\PT}\!\!\!\! \D^{3}Z_1\wedge \D^{3}Z_2 \wedge Z_1^{\alpha_I}\,Z_2^{\beta_I} \,g_{\beta_I}(Z_1)\wedge g_{\alpha_I}(Z_2)\,,
\ee
for $|I|=s-1$. The first term is local on twistor space, with $\dbar$ the natural (integrable) complex structure on $\CP^3$ and $\D^{3}Z$ the weight $+4$ holomorphic measure on $\PT$. Using the incidence relations, this volume form can be split into the weight $+2$ holomorphic measure on the $\CP^1$ fibres of $\PT$ and the ASD 2-forms on $\M$:
\be\label{PTvol}
\D^{3}Z = \sigma_{A}\sigma_{B}\, \d^{2}x^{AB}\wedge\D\sigma\,,
\ee
with $\D\sigma:=\sigma_{A}\d\sigma^{A}$. The second term in \eqref{linTCHS} is non-local, integrated over the fiber-wise product of twistor space with itself, $\PT\times_{\M}\PT\cong\M\times\CP^1\times\CP^1$. In particular, the twistors $Z_1$, $Z_2$ are functions of the \emph{same} $x\in\M$ but are located and \emph{different} points on the fibre: $Z_1=Z(x,\sigma_1)$, $Z_2=Z(x,\sigma_2)$. Equivalently, the measure for this non-local term may be written as
\be\label{nlmeas}
\int\limits_{\PT\times_{\M}\PT}\!\!\!\! \D^{3}Z_1\wedge\D^{3}Z_2\,(\cdots) = \int\limits_{\M\times\CP^1\times\CP^1}\!\!\!\!\d^{4}x\wedge\D\sigma_1\wedge\D\sigma_2\,(\sigma_1\,\sigma_2)^{2}\,(\cdots),
\ee 
to manifest the structure of the fibre-wise product. Here and subsequently, round brackets $(\sigma_1\,\sigma_2):=\sigma_{1A}\sigma_2^{\prime A}$ denotes the $\SL(2,\C)$-invariant inner product on the homogeneous coordinates of $\CP^1$.

To see that \eqref{linTCHS} is equivalent to \eqref{linCHS5}, consider the twistor space equations of motion:
\begin{align}
\label{linTeom1}
\dbar f^{\alpha_I} & =\varepsilon^{2}\,Z^{\beta_I}\,\d^{2}x^{CD} \int_{X}Z^{\prime\alpha_I}\, \sigma'_{C}\sigma'_{D}\,\D\sigma'\,g_{\beta_I}(Z')\,, \\
\label{linTeom2}
\dbar g_{\alpha_I}& =0\,.
\end{align}
Using the Penrose transform for vector- or form-valued $(0,1)$-forms on twistor space~\cite{Mason:1987,Mason:1990}, it can be shown that the twistorial equations \eqref{linTeom1}--\eqref{linTeom2} are equal to the space-time equations of motion \eqref{linCHS5}~\cite{Haehnel:2016mlb}. Of particular importance are the relationships ($|I| = s-1$)
\begin{equation*}
\dbar f^{\alpha_I} = \varepsilon^2 \left(0, \sigma^{B(s-1)}\, \Psi_{B(s-1)A(s-1)CD}\,\d^{2}x^{CD}\right)\,,
\end{equation*}
\begin{equation*}  
Z^{\beta_I}\,\d^{2}x^{CD} \int_{X}Z^{\alpha_I}\, \sigma'_{C}\sigma'_{D}\,\D\sigma'\,g_{\beta_I}(Z')=\left(0,\sigma^{B(s-1)}\, G_{B(s-1)A(s-1)CD}\,\d^{2}x^{CD}\right)\,,
\end{equation*} 
with equation \eqref{linTeom2} imposing $\nabla^{A(s)A'(s)}G_{A(s)B(s)}=0$ on the space-time spinor field. 

\medskip

Formulating the free CHS theory in twistor space suggests a natural way to describe the fully interacting model, following the examples of twistor actions for Yang-Mills and conformal gravity~\cite{Mason:2005zm,Adamo:2013tja}. The relevant data on twistor space is a copy of the pair \eqref{tf} for all $s\geq1$, now on a generic $\CPT$. The collection of all $f^{\alpha_I}$ defines a holomorphic structure on the \emph{infinite jet bundle} of $\CPT$ as a deformation of the trivial structure
\footnote{This jet space formulation should essentially be connected to the unfolded approach e.g. \cite{Vasiliev:1988sa, Vasiliev:1988xc}
and making this clearer is likely to be important. } 
\be\label{holos}
\dbar_{f}:=\dbar+f =\dbar+\sum_{|I|=0}^{\infty} f^{\alpha_I}
%				\frac{\partial^{|I|}}{\partial \Zt^{\alpha_I}}\,.
				\partial_{\alpha_I}\,.
\ee
This holomorphic structure is integrable if and only if the tensor~\cite{Haehnel:2016mlb}
\be\label{genNij}
N^{\alpha_I}:=\dbar f^{\alpha_I} + \sum_{|J|=0}^{|I|}\sum_{|K|=0}^{\infty} \binom{|J|+|K|}{|J|} f^{\beta_{K}(\alpha_{J}}\wedge\partial_{\beta_{K}} f^{\alpha_{I-J})}\,,
\ee
vanishes for every $|I|\geq0$. This should be equivalent to the vanishing of a higher-spin Nijenhuis tensor by a generalisation of the Newlander-Nirenberg theorem. (The multi-index $I-J$ corresponds to the complement of $J$ in $I$.) Note that this structure can be consistently truncated to the $s=1$ or $s=2$ cases, where it coincides with the $(0,2)$-curvature of an abelian gauge theory on $\CPT$ or the ordinary Nijenhuis tensor, respectively. However, for $|I|\geq2$, $N^{\alpha_I}$ contains source terms from all lower spin fields; thus, to extend consistently beyond spin two, the full tower of all higher spins must be included.

It follows naturally from the $s=1,2$ cases that the condition $N^{\alpha_{I}}=0$ should correspond to self-duality of the spin $s=|I|+1$ CHS gauge field. To impose this condition dynamically, each $N^{\alpha_I}$ is coupled to a corresponding Lagrange multiplier $g_{\alpha_I}$, giving a twistor action for SD interacting CHS theory:%
\footnote{To define $N^{\alpha_I}$ on $\CPT$ the deformation $f$ has to be promoted from a field on $\PT$ to a field on $\CPT$.}
\be\label{SDTA}
S_{\mathrm{SD}}[f,g]=\int_{\CPT} \D^{3}\Zt\wedge \sum_{|I|=0}^{\infty}g_{\alpha_{I}}\wedge N^{\alpha_I}\,.
\ee   
The resulting field equations are simply
\be\label{SDTeom}
N^{\alpha_{I}}=0\,, \qquad \dbar_{f}g_{\alpha_{I}}=0\,,
\ee
for all $|I|=0,\ldots,\infty$. These equations should be interpreted as the non-linear SD constraint (\textit{i.e.}, $\Psi_{A(s)B(s)}=0$) and the on-shell equation for a linearized ASD field propagating on this SD background (\textit{i.e.}, $\nabla^{A(s)A'(s)}G_{A(s)B(s)}=0$), respectively. 

Let us emphasize that there is currently no known formulation of interacting CHS theory on space-time in terms of Weyl curvatures. The interacting twistor action \eqref{SDTA} certainly suggests that such a formulation should exist, and that it will have a perturbative expansion around the SD sector (just like the linear theories). A \emph{proof} that this twistor action (along with all ASD interactions -- see below) is equivalent to the space-time CHS theory has yet to be provided, but we take the naturality of this action, along with the facts that it has the correct spectrum and linear truncation, to be strong evidence in its favour. In any case, \eqref{SDTA} certainly provides a well-defined interacting theory written in twistor space which, as we will see, has sensible scattering amplitudes.

\medskip

In~\cite{Haehnel:2016mlb} a non-linear extension for the ASD interaction term was also conjectured. This was quadratic in the negative helicity CHS fields and defined on a non-linear, self-dual CHS background. While~\cite{Haehnel:2016mlb} suggested possible geometric formulations of these interactions for arbitrary $s$, we only know a concrete formulation when the self-dual background is purely spin-two. Following~\cite{Adamo:2013tja}, the ASD interaction term for arbitrary negative helicity CHS fields on such a conformal gravity background is:
\be\label{ASDTA}
S^{(2)}_{\mathrm{int}}[f^{(2)},g]=\int\limits_{\CPT\times_{M}\CPT}\!\!\!\Omega_1\wedge\Omega_2\,\sum_{|I|=0}^{\infty}\Zt_1^{\alpha_{I}}\,\Zt_2^{\beta_{I}} \,g_{\beta_{I}}(\Zt_1)\, g_{\alpha_{I}}(\Zt_2)\,.
\ee
Here, $\CPT\times_{M}\CPT$ is the fibre-wise product of the curved twistor space associated with the spin-two holomorphic structure $\dbar_f = \dbar+f^{\alpha}\partial_\alpha $; $M$ is the associated four-dimensional SD space-time. The fibres $X\cong\CP^1$ are defined as the rational curves in $\CPT$ which are holomorphic with respect to this complex structure:
\be\label{hcurve}
\dbar_f \Zt^\alpha(x,\sigma) = 0
\qquad \Longrightarrow \qquad
\dbar \Zt^{\alpha}(x,\sigma)= f^{\alpha}(\Zt(x,\sigma))\,,
\ee
for fixed $x\in M$. The weight $+4$ $(3,0)$-form $\Omega$ is the top holomorphic form on this curved twistor space. 

This formulation differs slightly from the one given in~\cite{Haehnel:2016mlb}, and is chosen because it makes the perturbative expansion around the SD spin two background more straightforward. Although we do not use the approach of~\cite{Haehnel:2016mlb} in this paper, we expect the two approaches to be equivalent. Note that \eqref{ASDTA} reduces, term-by-term, to the free CHS actions \eqref{linTCHS} when $f^{\alpha}$ is set to zero.

The full interacting CHS theory is thus given on twistor space by:
\be\label{intTA}
S[f,g]=S_{\mathrm{SD}}[f,g]-\frac{\varepsilon^2}{2}\,S^{(2)}_{\mathrm{int}}[f^{(2)},g]-\frac{\varepsilon^2}{2}\sum_{s=2}^{\infty}S^{(s)}_{\mathrm{int}}[f,g]\,,
\ee
where the infinite sum of non-local interaction terms is defined conjecturally in~\cite{Haehnel:2016mlb}. We will here only consider the contributions to the action from $S_{\mathrm{SD}}$ and $S_{\mathrm{int}}^{(2)}$.

%%%%%%%%%%%%%%%%%%%%%%%%%%%%%%%%%%%%%%%%%%%%%%%%%%%%%%%

\subsection{A unitary subsector}

The CHS equations of motion on space-time \eqref{chseom1} make it clear that unitary (\textit{i.e.}, two-derivative) spin-$s$ fields, whose generalised Weyl spinors obey \eqref{zrm1}, are a consistent subsector of solutions. Twistor geometry provides an elegant way to restrict CHS theory to its unitary subsector. Thus far, the data on $\CPT$ preserves conformal invariance: the generalized complex structure $\dbar_{f}$ encodes only the self-duality of the CHS fields, and the Lagrange multipliers $g_{\alpha_I}$ represent solutions to the higher-derivative equations of motion via the Penrose transform.

Clearly, some additional structure is needed on twistor space in order to break conformal invariance. This is provided by specifying an \emph{infinity twistor}, $I_{\alpha\beta}$, which fixes the conformal factor of the space-time metric. The most useful choice of infinity twistor will be one corresponding to a vacuum Einstein conformal structure. If $\Lambda$ denotes the constant scalar curvature of such a vacuum, then the infinity twistor and its inverse are:
\be\label{infty}
I_{\alpha\beta}=\left(\begin{array}{cc}
                       \Lambda \epsilon_{A'B'} & 0 \\
                       0 & \epsilon^{AB}
                      \end{array}\right)\,, \qquad                     
I^{\alpha\beta}=\left(\begin{array}{cc}
                       \epsilon^{A'B'} & 0 \\
                       0 & \Lambda \epsilon_{AB}
                      \end{array}\right)\,.
\ee
These obey $I_{\alpha\beta}I^{\beta\gamma}=\Lambda\delta^{\gamma}_{\alpha}$, so in the flat space limit $\Lambda\rightarrow 0$ the infinity twistor becomes degenerate. Contractions with the infinity twistors are denoted using angle or square brackets,
\begin{equation*}
 I_{\alpha\beta}\,A^{\alpha}\,B^{\beta}:=\la A,B\ra\,, \qquad I^{\alpha\beta}\, C_{\alpha}\,D_{\beta}:=[C,D]\,,
\end{equation*}
as usual.

Following the example of $s=2$ conformal gravity~\cite{Adamo:2012nn}, the infinity twistor can be used to restrict to the unitary subsector for CHS twistor data~\cite{Haehnel:2016mlb}. In particular, for each $|I|=1,\ldots,\infty$ this restriction is defined by
\be\label{uss1}
f^{\alpha_{I}}=I^{\beta_I \alpha_I}\partial_{\beta_I}\,h^{(s)}\,, \qquad g_{\alpha_I}=I_{\alpha_I \beta_I}\,\Zt^{\beta_I}\,\tilde{h}^{(s)}\,,
\ee
where multi-indices on infinity twistors stand for $I_{\alpha_I\beta_I}:=I_{\alpha_1 \beta_1}\cdots I_{\alpha_{s-1}\beta_{s-1}}$, and so forth. By the anti-symmetry of the infinity twistor, this restriction is compatible with the conditions \eqref{gt}. 

For $|I|=s-1$ there are only two degrees of freedom on twistor space, parametrized by
\be\label{uss2}
h^{(s)}\in\Omega^{0,1}(\CPT,\cO(2s-2))\,, \qquad \tilde{h}^{(s)}\in\Omega^{0,1}(\CPT,\cO(-2s-2))\,.
\ee
When these are on-shell (\textit{i.e.}, cohomology classes) they produce helicity $\pm s$ solutions to the zero-rest-mass equations \eqref{zrm1} on space-time by the usual Penrose transform. The spectrum in this unitary subsector agrees with that found by Fronsdal~\cite{Fronsdal:1978rb} for the massless limit of Hagen-Singh theory~\cite{Singh:1974qz}.

%%%%%%%%%%%%%%%%%%%%%%%%%%%%%%%%%%%%%%%%%%%%%%%%%%%%%
%%%%%%%%%%%%%%%%%%%%%%%%%%%%%%%%%%%%%%%%%%%%%%%%%%%%%

\section{CHS Scattering Amplitudes from the Twistor Action}
\label{CHSamp}

The simplest semi-classical observables in any field theory with an action description are tree-level scattering amplitudes. For $s>2$ there is surprisingly little known about the tree-level S-matrix of CHS theory; in this section we use the twistor action to compute all 3-point tree amplitudes with two positive helicity external particles (\textit{i.e.}, $\overline{\mbox{MHV}}$ amplitudes) and all $n$-point tree amplitudes for two helicity $-s$ states and $(n-2)$ helicity $+2$ states (\textit{i.e.}, MHV amplitudes). The advantage of the twistor framework is that these amplitudes are generated by the SD and non-local parts of the twistor action, respectively. In the space-time formulation of CHS theory, these amplitudes would receive multiple Feynman diagram contributions involving complicated interaction vertices and space-time propagators (\textit{c.f.}, \cite{Joung:2015eny,Beccaria:2016syk}).

%%%%%%%%%%%%%%%%%%%%%%%%%%%%%%%%%%%%%%%%%%%%%%%%%%%%%

\subsection{Tree-level S-matrix \& external states}

In general, the definition of the S-matrix for massless particles with higher spin requires some subtlety. However, as we are only considering tree-level the definition of scattering amplitudes needed for our purposes can be fairly na\"ive. The semi-classical S-matrix for any theory with a Lagrangian description is encoded by a generating functional defined purely in terms of the classical action -- this is the basic content of the LSZ prescription. The external states of any scattering process are free fields which solve the linearized equations of motion on the scattering background. A $n$-point tree-level scattering amplitude is given by the piece of the generating functional which is multilinear in $n$ such on-shell external states. This perspective applies just as well to CHS theory formulated on twistor space, with the classical action given by \eqref{intTA} and a `flat' scattering background $\PT$. 

In four space-time dimensions, the tree-level S-matrix of any field theory which admits a perturbative expansion around the SD sector will posses an important structure: scattering amplitudes can be classified by their `MHV degree.' This means that the amplitudes are specified by on-shell four-momenta and a helicity label (rather than a polarization). Consistency and integrability of the SD sector imply that amplitudes for which fewer than two external states have negative helicity vanish; the `maximal helicity violating' -- or MHV -- amplitudes are those with two negative helicity external states and the rest positive helicity. At three points (for complexified kinematics), there is also the possibility of an $\overline{\mbox{MHV}}$ amplitude, with one negative helicity and two positive helicity external legs. We denote the $n$-point tree-level amplitude with $k$ negative helicity external states by $\cM_{n,k-2}$.

\medskip 

An advantage of the twistor formulation of CHS theory is that the generating functionals for $\overline{\mbox{MHV}}$ and MHV amplitudes are provided by the classical action itself. The SD portion of the action generates all 3-point amplitudes with a single negative helicity and two positive helicity CHS external states (of arbitrary spin). These are precisely the $\overline{\mbox{MHV}}$ amplitudes, packaged in the generating functional
\be\label{MHVbargf}
I^{\overline{\mathrm{MHV}}}=\sum_{|I|=0}^{\infty}\int_{\CPT}\D^{3}\Zt\wedge g_{\alpha_I}\wedge \sum_{|J|=0}^{|I|}\sum_{|K|=0}^{\infty}\binom{|J|+|K|}{|J|} f^{\beta_{K}(\alpha_{J}}\wedge\partial_{\beta_{K}} f^{\alpha_{I-J})}\,,
\ee
where $g_{\alpha_I}$ and $f^{\alpha_I}$ obey the linearized equations of motion: $\dbar g_{\alpha_I} = 0 = \dbar f^{\alpha_I}$.

All amplitudes with $(n-2)$ positive helicity conformal gravitons and two helicity $-s$ CHS fields are generated by the non-local term \eqref{ASDTA} in the twistor action. These constitute all the tree-level MHV amplitudes of CHS theory on a conformal gravity background, with generating functional \eqref{ASDTA}:
\be\label{MHVgf}
	%I^{\mathrm{MHV}}=\sum_{I=0}^{\infty}\,\int\limits_{\CPT\times_{M}\CPT}\!\!\!\Omega_{1}\wedge\Omega_{2}\,\Zt_{1}^{\alpha_{I}}\,\Zt_{2}^{\beta_{I}} \,g_{1\,\beta_{I}}(\Zt_{1})\, g_{2\,\alpha_{I}}(\Zt_{2})\,.
	I^{\mathrm{MHV}} = \int\limits_{\CPT\times_{M}\CPT}\!\!\!\Omega_1\wedge\Omega_2\,\sum_{|I|=0}^{\infty}\Zt_1^{\alpha_{I}}\,\Zt_2^{\beta_{I}} \,g_{\beta_{I}}(\Zt_1)\, g_{\alpha_{I}}(\Zt_2)\,.
\ee
The $n$-point amplitude is obtained from this functional by perturbatively expanding the integrand to order $(n-2)$ in $\Zt$ from solving \eqref{hcurve}, thus generating interactions with $f^{\alpha}$, which encodes the SD conformal gravity background, and evaluating the result on the flat fibre-wise product, $\PT\times_{\M}\PT$. Once more, on-shell external states obey $\dbar g_{\alpha_I} = 0 = \dbar f^{\alpha_I}$. 

More generally, any MHV amplitude can be seen as the amplitude for a negative helicity incoming state to flip helicity after crossing a SD background~\cite{Mason:2008jy,Adamo:2013tja,Adamo:2013cra}. In the case of \eqref{MHVgf}, this is the amplitude for a negative helicity spin-$s$ CHS field to flip helicity after crossing a SD conformal gravity background. The generating functional is quadratic in the negative helicity CHS states, defined on a non-linear, Bach-flat (in particular, self-dual) spin-two background. It is therefore natural to conjecture that \eqref{MHVgf} is equivalent to the SD part of quadratic covariant action for CHS fields on a conformal gravity background, whose existence was argued in~\cite{Grigoriev:2016bzl} (see also \cite{Nutma:2014pua} and \cite{Joung:2012qy, Joung:2016naf}). We hope to investigate this further in the future. 

\medskip

Let us briefly contrast the twistor space and `traditional' space-time perspectives on computing these $\overline{\mbox{MHV}}$ and MHV amplitudes. In the traditional picture, computing the three-point $\overline{\mbox{MHV}}$ amplitude requires computing the three point vertex for helicities $-s_{1}$, $s_{2}$ and $s_{3}$ from the interacting action. This interaction must then be evaluated on external states which obey the linearized CHS equations of motion. However, computing even these cubic vertices is non-trivial in the traditional approach, since the interacting CHS theory is defined as an induced theory.  On twistor space, one simply takes the $|I|=s_{1}-1$, $|J|+|K|=s_{2}-1$, and $|I|-|J|=s_{3}-1$ terms from \eqref{MHVbargf}.

This contrast with traditional, space-time based methods is even more drastic for the MHV amplitudes encoded by \eqref{MHVgf}. To compute the $n$-point amplitude in this class, one must determine the Feynman rules of the induced CHS action on space-time to capture all tree-level interactions between $(n-2)$ positive helicity conformal gravitons and two negative helicity spin $s$ CHS states. Since the induced action is non-polynomial in spin-two states (even the bare conformal gravity action is non-polynomial in the metric), this requires computing $n$-point vertices alongside the many contributions from diagrams containing propagators. Indeed, this is sufficiently complicated that only the 4-point calculation has been done explicitly~\cite{Joung:2015eny,Beccaria:2016syk}. The twistor action sidesteps this difficulty by rewriting CHS theory as a perturbative expansion around the SD sector: the MHV amplitudes are entirely captured by the non-local ASD interaction term, leading to the generating functional \eqref{MHVgf}.   

\medskip

To evaluate the amplitudes produced by the generating functions \eqref{MHVbargf} and \eqref{MHVgf}, explicit representations for the external states are needed. How to construct these representatives is not immediately obvious since $f^{\alpha_I}$ and $g_{\alpha_I}$ have free twistor indices. A particularly useful framework is provided by the \emph{helicity raising/lowering} formalism in twistor space~\cite{Penrose:1986ca}.\footnote{Our treatment here is valid on a conformally flat background; more generally, the equations will be modified by various correction terms involving the trace-free Ricci curvature.} 

Consider the $s=2$ example of conformal gravity; a negative helicity, on-shell conformal graviton is encoded by $g_{\alpha}\in H^{0,1}(\PT,\cO(-5))$. This field has the same homogeneity as $\psi\in H^{0,1}(\PT,\cO(-5))$, which describes a helicity $-\frac{3}{2}$ Rarita-Schwinger field by the Penrose transform:
\be\label{RS1}
\Psi_{ABC}(x) = \int_{X}\lambda_{A}\lambda_{B}\lambda_{C}\,\D\lambda\wedge\psi|_{X}\,, \qquad \nabla^{AA'}\Psi_{ABC}=0\,,
\ee
where $\D\lambda = \epsilon^{AB}\lambda_{A}\d\lambda_{B}$ is the weight $+2$ holomorphic measure on $X\cong\CP^1$ in twistor space. The Penrose transform of $g_\alpha$ yields a space-time field with a twistor index
\be\label{PTg}
 G_{\alpha BCD}(x) = \int_{X}\lambda_{B}\lambda_{C}\lambda_{D}\,\D\lambda\wedge g_\alpha |_{X}\,, \qquad \nabla^{BB'} G_{\alpha BCD}=0\,,
\ee
which splits into primed and unprimed part $G_{\alpha BCD} = (\gamma_{A'BCD}, G^A{}_{BCD})$. The covariant derivative acting on a local twistor index gives on conformally flat background
\be\label{cge1}
 	\nabla^{B}{}_{B'} G^{A}{}_{BCD} = \im\, \epsilon^{BA}\,\gamma_{B'BCD} \,,
 	\qquad 
 	\nabla^{B}{}_{B'} \gamma_{A'BCD} = 0\,,
\ee
subject to $G^{A}{}_{ABC}=0$. These equations can be solved by choosing a fixed dual twistor $(\tilde{\beta}_{A'},\beta^{A})$ obeying, in a conformally flat background,
\be
\label{dte}
 {\nabla^{B}}_{B'}\,\beta^{A}=\im\,\epsilon^{BA}\,\tilde{\beta}_{B'}\,,
 \qquad
 {\nabla^{B}}_{B'}\,\tilde{\beta}_{A'}=\im\,\Lambda \epsilon_{B'A'}\,{\beta}^{B}\,.
\ee
Using these components, we define the space-time spinors
\be\label{RS2}
G^{A}{}_{BCD}=\beta^{A}\,\Psi_{BCD}\,, \qquad \gamma_{A'BCD}=\tilde{\beta}_{A'}\,\Psi_{BCD}\,.
\ee
By virtue of the dual twistor equation, these spinors satisfy \eqref{cge1} respectively. 
%Even though \eqref{PTg} is conformally invariant, \eqref{cge1} is not (the Ricci tensor is not conformally invariant) and we have to account for that by considering the conformal class of \eqref{RS2}.
These equations \eqref{dte} are solved in flat-space, $\Lambda=0$,  by taking $\beta^{A}= \beta_0^{A}+\im x^{AA'}\tilde{\beta}_{0}{}_{A'}$ for some constant spinors $\beta^{A}_{0}, \tilde{\beta}_{0}{}_{A'}$. 
Solutions in arbitrary conformally flat geometries can be found by making the  conformal transformation  $g_{ab} \to \Omega^2 g_{ab}$\,. This transformation acts on the spinor components as 
\be 
	\beta^A \to \beta^A = \beta_0^A + \im x^{AA'} \tilde \beta_{0}{}_{A'}
	\,, \qquad 
	\tilde \beta_{A'} \to \tilde \beta_{0}{}_{A'} + \im \Upsilon_{AA'} \beta^A\,,
\ee
where $\Upsilon_{AA'} = \Omega^{-1} \nabla_{AA'} \Omega$. Note that only the primed component transforms.

To obtain a general solution to \eqref{cge1}, we also have to take into account the solution for the homogeneous case when $\tilde{\beta}_{A'}=0$\,. Let $\xi^\alpha=(\tilde{\xi}^{A'},\xi_{A})$ be a fixed twistor satisfying ${\nabla_A}^{A'} \tilde\xi^{B'} = -\im \epsilon^{A'B'} \xi_A$, then the full solution is given by:
\be\label{RS3}
G^{A}{}_{BCD}=\tilde{\xi}^{B'}{\nabla^{A}}_{B'}\Psi_{BCD}-4\im\epsilon^{AE}\,\Psi_{(BCD}\xi_{E)} +\beta^{A}\,\Psi_{BCD}\,.
\ee
It is straightforward to see that the two equations \eqref{cge1} imply the Bach equation
\begin{equation*}
\nabla_{AA'}\nabla^{B}{}_{B'}G^A{}_{BCD}=0\,,
\end{equation*}
so the solution constructed in this way from the Rarita-Schwinger field encodes the ASD modes of conformal gravity. 
In order to write the solution \eqref{RS3} as $g_{\alpha}=B_{\alpha}\psi$ on twistor space for $\psi\in H^{0,1}(\PT,\cO(-5))$ we note that
\be
\int_X \lambda_{(B} \lambda_C \lambda_D \xi_{A)} \,\D\lambda \wedge \psi|_{X}
	= -\frac{1}{4}\int_X \lambda_{B} \lambda_C \lambda_D \lambda_{A} \xi_{E} \,\D\lambda \wedge \partial^E \psi|_{X}
\ee
since $\xi_E$ is independent of $\lambda$, such that
\be\label{HLO}
B_{\alpha}=\left(\tilde{\beta}_{A'}  \,,\:\: i \lambda^{A} {\xi}^{\delta}\partial_\delta+\beta^{A} \right) \,.
\ee
Since $B_{\alpha}$ is holomorphic and homogeneous, it follows that $g_{\alpha}\in H^{0,1}(\PT,T^{*}_{\PT}(-5))$, as desired. The degrees of freedom of conformal gravity are now manifest: in the flat space limit $(\tilde{\xi}^{A'},\xi_{A})$ parametrizes the helicity $-2$ conformal graviton, $\tilde{\beta}_{A'} $ parametrizes the conformal ghost, and $\beta^{A}$ pa\-ra\-me\-trizes a conformal spin one state.

Similarly, $f^{\alpha}\in H^{0,1}(\PT,T_{\PT}(1))$ can be constructed from a helicity $+\frac{3}{2}$ representative $\tilde{\psi}\in H^{0,1}(\PT,\cO(1))$ and a helicity raising operator $A^{\alpha}$:
\be\label{HRO}
A^{\alpha}=\left(\zeta_{\gamma}Z^\gamma\, \frac{\partial}{\partial\mu_{A'}} +\tilde{\alpha}^{A'}\,, \:\: \alpha_{A} \right)\,, 
\ee
where $\zeta_\alpha=(\tilde{\zeta}_{A'},\zeta^{A})$ is a dual twistor parametrizing the helicity $+2$ conformal graviton, $\tilde{\alpha}^{A'}$ is a spinor parametrizing a conformal spin one state, and $\alpha_{A}$ parametrizes the conformal ghost in analogous fashion as above. Since this $A^{\alpha}$ is holomorphic and homogeneous, $f^{\alpha}=A^{\alpha}\tilde{\psi}$ encodes the appropriate on-shell information in twistor space.

This procedure of helicity raising or lowering can be applied repeatedly to build on-shell representatives in twistor space for arbitrary spin CHS fields. For $|I|=s-1$, the twistor data  $g_{\alpha_I}$, $f^{\alpha_{I}}$ is constructed as
\be\label{HRL}
g_{\alpha_I}=B_{\alpha_I}\,\psi^{(-3-s)}\,, \qquad f^{\alpha_I}=A^{\alpha_I}\,\tilde{\psi}^{(s-1)}\,,
\ee
where the lowering/raising operators are products of copies of \eqref{HLO} or \eqref{HRO}, and the fields $\psi^{(-3-s)}\in H^{0,1}(\PT,\cO(-3-s))$, $\tilde{\psi}^{(s-1)}\in H^{0,1}(\PT,\cO(s-1))$ are representatives for zero-rest-mass fields of helicity $\mp\frac{s+1}{2}$, respectively. In standard space-time language, this means that positive and negative helicity CHS states of spin $s$ can be constructed from positive and negative helicity massless, two-derivative states of spin $\frac{s+1}{2}$.

\medskip

Of course, explicit representatives for $\psi^{(-3-s)}$ and $\tilde{\psi}^{(s-1)}$ are needed to obtain closed-form expressions for scattering amplitudes. A particularly convenient choice is provided by \emph{dual twistor} wavefunctions, which associate a fixed dual twistor $W_{\alpha}=(\tilde{\lambda}_{A'},\tilde{\mu}^{A})$ to each external field. The representatives take the form of plane-waves in twistor space, with this dual twistor serving as the `momentum':
\be\label{dtwf1}
\psi^{(-3-s)}(Z;W)=\int\d t\,t^{s+2}\,\e^{t\,W\cdot Z}\,, \qquad \tilde{\psi}^{(s-1)}(Z;W)=\int \frac{\d t}{t^{s}}\,\e^{t\,W\cdot Z}\,.
\ee
The integrals over the scaling parameter $t$ ensure that the expressions have the correct homogeneity on twistor space. A major advantage of working with dual twistor wavefunctions is that they render differential operators in twistor space as algebraic expressions in the dual twistors. Any amplitude expression in terms of dual twistors can then be transformed into an expression on momentum space by means of the half-Fourier transform. Note that since $g_{\alpha_I}$ and $f^{\alpha_I}$ are form-valued, the exponentials $e^{t W\cdot Z}$ are form-valued as well and yield signs in concrete computations when commuting them. To lighten the notation, we will treat any product $e^{t_i W_i\cdot Z} \cdot e^{t_j W_j\cdot Z}$ as an ordered product ($i<j$) and include the appropriate sign in the pre-factors accordingly. To summarize, arbitrary spin CHS states will be represented on twistor space by:
\be\label{dtwf2}
g_{\alpha_I}=B_{\alpha_I}\int \d t\,t^{|I|+3}\,\e^{t\,W\cdot Z}\,, \qquad f^{\alpha_I}=A^{\alpha_I}\int\frac{\d t}{t^{|I|+1}}\,\e^{t\,W\cdot Z}\,,
\ee
where $|I|=s-1$.

%%%%%%%%%%%%%%%%%%%%%%%%%%%%%%%%%%%%%%%%%%%%%%%%%%%%%

\subsection{All \texorpdfstring{$\overline{\mbox{MHV}}$}{anti-MHV} amplitudes}

Consider the contribution to the generating function \eqref{MHVbargf} with arbitrary but fixed external spins $s_1$, $s_2$, $s_3$. One state, say that with spin $s_1$, has negative helicity while the other two have positive helicity. When evaluated with on-shell states, this contribution gives the 3-point $\overline{\mbox{MHV}}$ amplitude:
\begin{multline}\label{MHVb1}
\cM_{3,-1}(-s_1, +s_2, +s_3)=\int\D^{3}Z\wedge g_{1\,\alpha_I}\wedge\left(\frac{(|J|+|K|)!}{|K|!\,|J|!}f_{2}^{\beta_{K}(\alpha_{J}}\wedge\partial_{\beta_K}f_{3}^{\alpha_{I-J})} \right.\\
\left. + \frac{(|I|-|J|)!}{|K|!\,(|I|-|J|-|K|)!}f_{3}^{\beta_{K}(\alpha_{I-J-K}}\wedge\partial_{\beta_K} f_{2}^{\alpha_{J+K})}\right)\,,
\end{multline}
where $|I|=s_1 -1$, $|J|+|K|=s_2 -1$, and $|I|-|J|=s_{3}-1$. Plugging in the on-shell states \eqref{dtwf2}, a bit of algebra shows that this expression is equal to
\begin{multline*}
\cN^{(s_1,s_2,s_3)}\int\frac{\d^{4}Z}{\mathrm{vol}\;\C^{*}}\frac{\d t_{1}\,\d t_{2}\,\d t_{3}}{t_{2}^{|K|+|J|+1} t_{3}^{|I|-|J|+1}} t_{1}^{|I|+3}\,B_{1\,\alpha_I}\,\times \\
\left((s_2-1)!\,(s_1-s_2)!\,t_3^{|K|}(A_2\cdot W_3)^{|K|} A_{2}^{\alpha_J} A_{3}^{\alpha_{I-J}} \right.\\
\left.+(-1)^{|K|+1}(s_3-1)!\,(s_1-s_3)!\,t_2^{|K|}(A_3\cdot W_2)^{|K|}A_{2}^{\alpha_{J+K}}A_{3}^{\alpha_{I-J-K}}\right)\,\e^{\sum_{i=1}^{3}t_i W_i\cdot Z}\,,
\end{multline*}
where we have re-written the projective measure $\D^{3}Z$ on $\PT$ as a non-projective measure on $\C^4$ modulo a quotient by $\C^{*}$-scalings. The overall normalisation constant is
\begin{equation*}
 \cN^{(s_1,s_2,s_3)}:=\frac{1}{(s_2+s_3-s_1-1)!\,(s_1-s_3)!\,(s_1-s_2)!}\,,
\end{equation*}
in terms of the external spins.

Using the vol $\C^{*}$ to fix $t_1=1$ and then performing the $\d^{4}Z$ integrals leaves
\begin{multline*}
\cN^{(s_1,s_2,s_3)}\int\frac{\d t_{2}\,\d t_{3}}{t_{2}^{s_2} t_{3}^{s_3}}\left((s_2-1)!\,(s_1-s_2)!\,(B_1\cdot A_2)^{|J|}(B_1\cdot A_3)^{|I|-|J|}(A_2\cdot W_1)^{|K|}\right. \\
\left. +(-1)^{|K|+1}(s_3-1)!\,(s_1-s_3)!\,(B_1\cdot A_2)^{|J|+|K|}(B_1\cdot A_3)^{|I|-|J|-|K|}(A_3\cdot W_1)^{|K|}\right) \\
\times \delta^{4}\!\left(W_1 +t_2 W_2 +t_3 W_3\right)\,
\end{multline*}
where the delta function and gauge conditions $A_i\cdot W_i=0$ have been used to eliminate some powers of $t_3$ and $t_2$ from each term respectively. At this point, the delta functions can be rearranged as
\begin{equation*}
\delta^{4}\!\left(W_1 +t_2 W_2 +t_3 W_3\right)=\frac{\delta^{2}(\tilde{\mu}_1+t_2\tilde{\mu}_2+t_3\tilde{\mu}_3)}{[23]}\,\delta\!\left(t_2 -\frac{[31]}{[23]}\right)\,\delta\!\left(t_{3}-\frac{[12]}{[23]}\right)\,,
\end{equation*}
where $[i\,j]:=\epsilon_{A'B'}\tilde{\lambda}^{A'}_{i}\tilde{\lambda}^{B'}_{j}$. The $\d t_2$, $\d t_3$ integrals are performed explicitly against these delta functions to give:
\begin{multline}\label{MHVb2}
\cM_{3,-1}(-s_1,+s_2,+s_3)\\
=\cN^{(s_1,s_2,s_3)}\left((s_2-1)!\,(s_1-s_2)!\,(B_1\cdot A_2)^{s_{1}-s_{3}}(B_1\cdot A_3)^{s_{3}-1}(A_2\cdot W_1)^{s_{23|1}}\right. \\
\;\, +\left. (-1)^{s_{23|1}+1}(s_3-1)!\,(s_1-s_3)!\,(B_1\cdot A_2)^{s_{2}-1}(B_1\cdot A_3)^{s_{1}-s_{2}}(A_3\cdot W_1)^{s_{23|1}}\right) \\
\times\frac{[23]^{s_2+s_3+1}}{[12]^{s_3}[31]^{s_2}}\, \delta^{2}\!\left([23]\tilde{\mu}_{1}+[31]\tilde{\mu}_{2}+[12]\tilde{\mu}_{3}\right)\,,
\end{multline}
using the shorthand
\be\label{scomb}
s_{ij|k}:=s_{i}+s_{j}-s_{k}-1\,.
\ee
In this expression, the helicity raising/lowering operators $A_{i}, B_{i}$ should be thought of as differential operators acting on everything to the right. Dependence on the $\tilde{\mu}$ variables can be removed via the half-Fourier transform
\be\label{HFT}
\cM_{3,-1}(\{\lambda_{i},\tilde{\lambda}_{i}\}):=\int \cM_{3,-1}(\{\tilde{\mu}_{i},\tilde{\lambda}_{i}\})\,\prod_{i=1}^{3}\D\tilde{\mu}_{i}\,\e^{\la\tilde{\mu}_{i}\lambda_{i}\ra}\,,
\ee
if desired.

Although the formula \eqref{MHVb2} for a general $\overline{\mbox{MHV}}$ amplitude may seem a bit unwieldy, it simplifies considerably in certain spin sectors. For instance, if all the external spins are equal $(s_1=s_2=s_3)$ then the amplitude is
\begin{multline}\label{MHVb3}
\cM_{3,-1}(-s,+s,+s)=\left((A_{3}\cdot B_{1})^{s-1}(A_{2}\cdot W_1)^{s-1}+(-1)^{s}(A_{2}\cdot B_{1})^{s-1} (A_{3}\cdot W_{1})^{s-1}\right) \\
\times \frac{[23]^{2s+1}}{[12]^{s}[31]^{s}}\,\delta^{2}\!\left([23]\tilde{\mu}_{1}+[31]\tilde{\mu}_{2}+[12]\tilde{\mu}_{3}\right)\,.
\end{multline}
As we will see, even further simplification occurs when the $A_i$ and $B_i$ are chosen to encode the unitary subsector of the CHS theory.

%%%%%%%%%%%%%%%%%%%%%%%%%%%%%%%%%%%%%%%%%%%%%%%%%%%%%

\subsection{MHV amplitudes on a conformal gravity background}

To compute the $n$-point MHV amplitude with $(n-2)$ positive helicity spin two states and two negative helicity spin $s$ states, we must perturbatively expand the generating functional \eqref{MHVgf}. This expansion is operationalized using the same techniques as in~\cite{Adamo:2013tja}. On the SD background $M$ composed of the helicity $+2$ states, the fibre-wise product $\CPT\times_{M}\CPT$ is governed by \eqref{hcurve}, which can be re-written as an integral equation:
\be\label{hcurve2}
\cZ^{\alpha}(x,\sigma) = X^{\alpha\,A}\,\sigma_{A}+\dbar^{-1}|_{X} f^{\alpha}(\cZ(x,\sigma))\,.
\ee
Here, $X^{\alpha\,A}$ parametrize the homogenous solution and $\dbar^{-1}|_{X}$ is the inverse of the $\dbar$-operator restricted to the holomorphic curve labeled by $x\in M$. Since $f^{\alpha}$ is homogeneous of weight $+1$ on twistor space, there is an ambiguity in the definition of $\dbar^{-1} f^{\alpha}$ which can be fixed by requiring $\dbar^{-1}_{X}$ to have a second-order zero at some fixed point $\xi\in\CP^1$. The choice of $\xi$ is entirely arbitrary, constituting a `gauge' in twistor space. 

Having made this choice, \eqref{hcurve2} can be expanded perturbatively around the homogenous solution as
\be\label{hcurve3}
\cZ^{\alpha}(x,\sigma)=X^{\alpha\,A}\,\sigma_{A} + \frac{1}{2\pi\im}\int_{\CP^1}\frac{\D\sigma '}{(\sigma\sigma')}\frac{(\xi\sigma)^2}{(\xi\sigma')^{2}}\,f^{\alpha}(X\cdot\sigma') \,+\cdots\,,
\ee   
where the $+\cdots$ terms are higher-order terms in the expansion arising from expanding $f^\alpha(\cZ(x,\sigma'))$ around $Z(x,\sigma')=X\cdot \sigma'$. The idea is to act with this expansion iteratively in the generating functional $(n-2)$ times. The first-order contribution is given by summing over all ways of shifting
\be\label{pertexp}
\cZ^{\alpha}(x,\sigma)\rightarrow  \frac{1}{2\pi\im}\int_{\CP^1}\frac{\D\sigma '}{(\sigma\sigma')}\frac{(\xi\sigma)^2}{(\xi\sigma')^{2}}\,f^{\alpha}(Z(x,\sigma'))\,,
\ee
in \eqref{MHVgf}. At second order, we sum over all the ways of similarly shifting the first-order contribution, and so on.

\medskip

It will be useful to rewrite the generating functional \eqref{MHVgf} in a slightly different way:
\begin{multline}\label{MHV1}
I^{\mathrm{MHV}}=\sum_{|I|=0}^{\infty}\,\int\limits_{M\times\CP^{1}\times\CP^{1}}\!\!\!\frac{\d^{8} X}{\mathrm{vol}\,\GL(2,\C)}\,(1\,2)^{2}\,\D\sigma_{1}\,\D\sigma_{2}\,\cZ^{\alpha_{I}}(x,\sigma_1)\,\cZ^{\beta_{I}}(x,\sigma_2) \\
g_{1\beta_{I}}(\cZ(x,\sigma_1))\, g_{2\alpha_{I}}(\cZ(x,\sigma_2))\,,
\end{multline}
where the holomorphic measure on the fibre-wise product has been converted into a measure on $M\times\CP^{1}\times\CP^{1}$, with $\d^{8}X\,(\mathrm{vol}\,\GL(2,\C))^{-1}$ the measure on the space of holomorphic curves in twistor space. The first iteration of the perturbative expansion can act either at $\cZ(x,\sigma_1)$ or $\cZ(x,\sigma_2)$, and in each case this action can be either in the explicit powers of $\cZ^{\alpha}$ in \eqref{MHV1} or in the wavefunctions $g_{1,2\alpha_I}$. In the latter case, the expansion takes the form of a derivation:
\be\label{pewf}
g_{i\alpha_I}(\cZ(x,\sigma_i))\rightarrow\frac{1}{2\pi\im}\int_{\CP^1}\frac{\D\sigma '}{(i\,\sigma')}\frac{(\xi\,i)^2}{(\xi\sigma')^{2}}\,f^{\beta}(Z(x,\sigma'))\,\frac{\partial }{\partial Z^{\beta}(x,\sigma_i)}g_{i\alpha_I}(Z(x,\sigma_i))\,.
\ee  
All subsequent iterations act similarly: the perturbation is either at an explicit insertion of $\cZ(x,\sigma_i)$ or at one of the wavefunctions, which now include $f^\alpha$ insertions from previous iterations.

After iterating $(n-2)$ times, the expansion is equivalent to summing over all $n$-point Feynman tree diagrams on the $\CP^{1}$ parametrized by $X^{\alpha A}$ and rooted at the locations $\sigma_{1},\sigma_{2}\in\CP^{1}$. This means that the matrix tree theorem can be used to organize the expansion of the generating functional in terms of weighted determinants, in much the same way that it can be applied in the context of tree amplitudes of Einstein gravity~\cite{Feng:2012sy,Adamo:2012xe}. With dual twistor wavefunctions \eqref{dtwf2}, the generating functional becomes
\begin{multline}\label{MHV2}
I^{\mathrm{MHV}}=\sum_{|I|=0}^{\infty}\,\int\frac{\d^{8} X}{\mathrm{vol}\,\GL(2,\C)}\,(1\,2)^{2}\,\D\sigma_{1}\,\D\sigma_{2}\,(B_{1}\cdot \cZ(x,\sigma_2))^{|I|}(B_{2}\cdot \cZ(x,\sigma_{1}))^{|I|} \\
\d t_{1}\,t_{1}^{|I|+3}\,\d t_{2}\,t_{2}^{|I|+3}\,\exp\left(t_{1}W_{1}\cdot \cZ(x,\sigma_1)+t_{2}W_{2}\cdot \cZ(x,\sigma_{2})\right)\,.
\end{multline}
At this point, we take an arbitrary but fixed contribution to perform the expansion, with the two negative helicity states of spin $s$.

The basic object for applying the matrix tree theorem is a weighted Laplacian matrix, $\sM$, encoding the action of the perturbative expansion. This is a $n\times n$ matrix with entries
\be\label{Mmat}
\sM_{ij}= (-1)^{|i-j|} t_{j}\,A_{i}\cdot W_{j}\,\frac{\sqrt{\D\sigma_{i}\D\sigma_{j}}}{(i\,j)}\,, \quad \mbox{ for } i\neq j\,,
\ee 
\begin{equation*}
\sM_{ii}=-\D\sigma_{i}\,\sum_{j\neq i} (-1)^{|i-j|} t_{j}\frac{A_{i}\cdot W_{j}}{(i\,j)}\frac{(\xi\,j)^2}{(\xi\,i)^2}\,,
\end{equation*}
following from the structure of \eqref{pewf}. The signs $(-1)^{|i-j|}$ arise from the ordering of the form-valued exponentials. The reduced determinant of this matrix, with rows and columns corresponding to $i=1,2$ removed, encodes the sum of all contributions arising from the expansion acting on wavefunctions. When the expansion acts at one of the explicit $\cZ(x,\sigma_i)$ insertions, we must remove an additional row and column from the determinant, insert
\be\label{mcont}
\m^{i}_{j}=(-1)^{|i-j|} \frac{\D\sigma_{j}}{(i\,j)}\frac{(\xi\,i)^2}{(\xi\,j)^2}\,,
\ee
and sum over all the possible ways of doing this.

Putting the pieces together gives the expression for the $n$-point MHV amplitude:
\begin{multline}\label{MHV3}
\cM_{n,0}=\int\frac{\d^{8}X}{\mathrm{vol}\;\GL(2,\C)}(1\,2)^{2}\,\D\sigma_{1}\,\D\sigma_{2}\Bigg[ (B_{1}\!\cdot\!X\!\cdot\!\sigma_2)^{s-1}(B_{2}\!\cdot\!X\!\cdot\!\sigma_{1})^{s-1}\,\left|\sM^{12}_{12}\right| \\
+(s-1)\sum_{i=3}^{n}\left(B_{1}\!\cdot\!A_{i}\,\m^{2}_{i}\,(B_{1}\!\cdot\!X\!\cdot\!\sigma_2))^{s-2}(B_{2}\!\cdot\!X\!\cdot\!\sigma_{1})^{s-1} + (1\leftrightarrow 2)\right) \left|\sM^{12i}_{12i}\right| \\
+\cdots+((s-1)!)^{2}\sum_{\substack{i_{1},\ldots,i_{s-1} \\ j_{1},\ldots,j_{s-1}}} \left(\prod_{a=1}^{s-1}B_{1}\!\cdot\!A_{i_a} B_{2}\!\cdot\!A_{j_a} \m^{2}_{i_a}\m^{1}_{j_a} \right) \left|\sM^{12i_{1}\cdots j_{s-1}}_{12i_{1}\cdots j_{s-1}}\right| \Bigg] \\
\times \e^{\im\cP\cdot X}\,(t_{1}t_{2})^{s+4}\prod_{k=1}^{n}\frac{\d t_{k}}{t_{k}^{2}}\,.
\end{multline}
In the final line, the generalized `momentum' $\cP_{\alpha A}$ is defined to be
\be\label{gmom}
\cP_{\alpha A}:=-\im\sum_{i=1}^{n}W_{i\,\alpha}\,\sigma_{i\,A}\,.
\ee
Though this formula may appear complicated at first glance, the structure of each contribution is quite simple. The first line contains all contributions to the amplitude from the perturbative expansion acting on external wavefunctions; each of the subsequent contributions is a sum over the ways in which the expansion can also act on explicit $\cZ(x,\sigma)$ insertions. This sum is exhausted by the contributions on the third line, where all explicit insertions of $\cZ(x,\sigma)$ have been eaten by the perturbative expansion. Of course, for $n<2s-2$ the expression for $\cM_{n,0}$ will terminate sooner. When $s=2$, \eqref{MHV3} agrees with the MHV amplitude obtained from the twistor action of conformal gravity~\cite{Adamo:2013tja}.

\medskip

An initial concern about this expression for the amplitude is its gauge invariance: it seems far from clear that \eqref{MHV3} is independent of the fixed point $\xi\in\CP^{1}$ used to define the perturbative expansion. Since \emph{any} choice of $\xi$ suffices in \eqref{hcurve3}, and this choice encodes no degrees of freedom relevant to the physical CHS theory on space-time, physical quantities (such as amplitudes) must be independent of $\xi$. This issue has arisen before in the context of gravitational amplitudes in Minkowski space~\cite{Cachazo:2012kg} or AdS$_4$~\cite{Adamo:2015ina}, as well as conformal gravity~\cite{Adamo:2013tja,Adamo:2013cra}, and identical steps can be followed to check gauge invariance for \eqref{MHV3}. A lengthy but straightforward calculation, proceeding from the definitions of $\sM$, $\m$, and the properties of determinants shows that
\be\label{ginv}
\frac{\partial\cM_{n,0}}{\partial\xi^{B}}=\int\frac{\d^{8}X}{\mathrm{vol}\;\GL(2,\C)}\,\frac{\partial\mathcal{V}^{\alpha A}_{B}}{\partial X^{\alpha A}} =0\,.
\ee
for $\mathcal{V}^{\alpha A}_{B}$ a smooth function with respect to $X^{\alpha A}$. In other words, gauge invariance follows by Stokes' theorem on the moduli space of holomorphic curves in twistor space.

%%%%%%%%%%%%%%%%%%%%%%%%%%%%%%%%%%%%%%%%%%%%%
%%%%%%%%%%%%%%%%%%%%%%%%%%%%%%%%%%%%%%%%%%%%%

\section{Restricting to the Unitary Subsector}
\label{Uamp}

So far, all amplitudes have been evaluated with totally general CHS external states. In this section, we restrict the external states to lie in the unitary subsector, defined on twistor space using the infinity twistor \eqref{uss1}. On space-time, this restriction corresponds to choosing external states which solve the two-derivative equations of motion within the more general CHS framework. In this subsector, the external degrees of freedom are encoded in the twistor wavefunctions $h^{(s)}$, $\tilde{h}^{(s)}$ of \eqref{uss2}. This restriction can be understood as choosing particular helicity raising and lowering operators involving the infinity twistors $I^{\alpha\beta}$, $I_{\alpha\beta}$ describing a conformally flat background geometry as in \eqref{infty}.
 Schematically we choose
\be
{A}^{\alpha}\rightarrow  I^{\beta\alpha}\partial_{\beta}~, ~~~{\rm and}~~~ B_{\alpha} \rightarrow I_{\alpha\beta}Z^{\beta}
\ee
while the remaining factors provide the appropriate scaling of the wavefunctions. 
The unitary wavefunctions have dual twistor representatives
\be\label{udtwf}
h^{(s)}=\int \frac{\d t}{t^{2s-1}}\,\e^{t\,W\cdot Z}\,, \qquad \tilde{h}^{(s)}=\int \d t\,t^{2s+1}\,\e^{t\,W\cdot Z}\,,
\ee
homogeneous of degree $2s-2$ and $-2s-2$, respectively. Since these wavefunctions solve the (two-derivative) zero-rest-mass equations on space-time, they are naturally based on plane waves with an on-shell (complex) four-momentum. Hence, it will be useful to compute amplitudes in terms of ordinary momentum eigenstates as well. These have a straightforward representation on momentum space~\cite{Adamo:2011pv}:
\be\label{momeig}
h^{(s)}=\int \frac{\d t}{t^{2s-1}}\,\bar{\delta}^{2}\left(t\,\lambda-k\right)\,\e^{t\,[\mu\tilde{k}]}\,, \quad \tilde{h}^{(s)}=\int \d t\,t^{2s+1}\,\bar{\delta}^{2}\left(t\,\lambda-k\right)\,\e^{t\,[\mu\tilde{k}]}\,,
\ee
in terms of the on-shell four-momentum $k^{\mu}\leftrightarrow k^{A}\tilde{k}^{A'}$.

It is worth noting that, in the unitary subsector, the helicity raising operator $A^{\alpha}$ is purely algebraic, while the helicity lowering operator $B_{\alpha}$ must still be treated as a differential operator. Indeed, at the level of components,
\be\label{uhos}
A^{\alpha}|_{\mathrm{unitary}}=I^{\alpha\beta}\,W_{\beta}\,, \qquad B_{\alpha}|_{\mathrm{unitary}}=I_{\alpha\beta}\,t^{-1}\frac{\partial}{\partial W_{\beta}}\,,
\ee
when acting on dual twistor wavefunctions. For each insertion of $A^{\alpha}$ or $B_{\alpha}$, the unitary subsector is given by a linear combination of the conformal graviton and ghost modes; the spin one conformal state is set to zero. In the Minkowski space limit, $\Lambda\rightarrow 0$, the ghost mode is killed and only a part of the conformal graviton mode survives.

\medskip

An amplitude with all external states in the unitary subsector is denoted with a tilde: $\widetilde{\cM}_{n,k}$ is the tree-level amplitude involving $k+1$ negative helicity, two-derivative external states and $n-k-1$ positive helicity, two-derivative external states. Although the external states are two-derivative zero-rest-mass fields, the amplitude is computed using the full interacting CHS theory. In other words, computing $\widetilde{\cM}_{n,k}$ with `traditional' methods would entail calculating Feynman rules for the interacting induced CHS action on space-time. For $k=0,1$, these amplitudes are given in twistor space by simply restricting the generating functionals \eqref{MHVbargf}, \eqref{MHVgf} to the unitary subsector.

For definite external spins $s_1$, $s_2$, and $s_3$, the 3-point $\overline{\mbox{MHV}}$ amplitude in the unitary subsector is given by:
\begin{multline}\label{umhvb1} 
\widetilde{\cM}_{3,-1}(-s_{1},+s_{2},+s_{3})=\int\D^{3}Z \wedge I_{\alpha_{I}\gamma_{I}}\,Z^{\gamma_{I}}\,\tilde{h}_{1} \\
\wedge\left[\frac{(|J|+|K|)!}{|J|!\,|K|!}\,I^{\alpha_{J}\delta_{J}}I^{\beta_{K}\mu_{K}}\,\partial_{\delta_J}\partial_{\mu_{K}}\, h_{2}\wedge I^{\alpha_{I-J}\nu_{I-J}}\,\partial_{\beta_K}\partial_{\nu_{I-J}}\,h_{3}\right. \\
\left.+\frac{(|I|-|J|)!}{(|I|-|J|-|K|)!\,K!}\,I^{\alpha_{I-J-K}\delta_{I-J-K}}I^{\beta_{K}\mu_{K}}\,\partial_{\delta_{I-J-K}}\partial_{\mu_K}\,h_{3}\wedge I^{\alpha_{J+K}\nu_{J+K}}\,\partial_{\beta_K}\partial_{\nu_{J+K}}\,h_{2}\right]\,,
\end{multline}
where $|I|=s_1-1$, $|J|+|K|=s_{2}-1$, and $|I|-|J|=s_{3}-1$. Recalling that $I^{\alpha\beta}I_{\beta\gamma}=\Lambda\delta^{\alpha}_{\gamma}$, this expression further simplifies to
\begin{multline}\label{umhvb2}
\Lambda^{|I|}\int\D^{3} Z\wedge Z^{\gamma_I}\,\tilde{h}_{1}\wedge\left[\frac{(|J|+|K|)!}{|J|!\,|K|!}\,I^{\beta_{K}\mu_{K}}\partial_{\mu_{K}}\partial_{(\gamma_{J}}\, h_{2}\wedge \partial_{\gamma_{I-J})}\partial_{\beta_K}\,h_{3}\right. \\
\left.+\frac{(|I|-|J|)!}{(|I|-|J|-|K|)!\,K!}\,I^{\beta_{K}\mu_{K}}\,\partial_{\mu_K}\partial_{(\gamma_{I-J-K}}\,h_{3}\wedge \partial_{\gamma_{J+K})}\partial_{\beta_K}\,h_{2}\right]\,.
\end{multline}
The overall spin-dependent power of the cosmological constant is an important feature of all CHS amplitudes restricted to the unitary subsector.

Now, the differential operator $Z\cdot\partial$ is just the Euler vector field on $\PT$; when acting on a homogeneous function $f$ of weight $\kappa$ it gives $Z\cdot\partial f=\kappa f$. Since all wavefunctions appearing in \eqref{umhvb2} have well-defined homogeneities, successive applications of the Euler vector results in:
\begin{multline}\label{umhvb3}
\widetilde{\cM}_{3,-1}(-s_{1},+s_{2},+s_{3})=\frac{\Lambda^{|I|}\,(2|J|+|K|)!\,(2|I|-2|J|-|K|)!}{|K|!\,|J|!\,(|I|-|J|-|K|)!}\int \D^{3}Z\wedge \tilde{h}_{1} \\
\wedge I^{\alpha_{K}\beta_{K}}\left(\partial_{\alpha_{K}}h_{2}\wedge\partial_{\beta_K}h_{3}+\partial_{\alpha_K}h_{3}\wedge\partial_{\beta_K}h_{2}\right) \\
=\Lambda^{s_{1}-1}\,\widetilde{\cN}^{(s_1,s_2,s_3)}\int_{\PT}\D^{3}Z\wedge\tilde{h}_{1}\wedge\left\{h_{2},h_{3}\right\}_{(s_{23|1}+1)}\,,
\end{multline}
where the unitary subsector normalization constant is:
\be\label{unorm}
\widetilde{\cN}^{(s_1,s_2,s_3)}:=\frac{1+(-1)^{s_{23|1}+1}}{(s_1-s_2)!\,(s_1-s_3)!}\,\frac{\Gamma(s_1+s_2-s_3)\, \Gamma(s_1-s_2+s_3)}{\Gamma(s_2+s_3-s_1)}\,,
\ee
and we have defined the bracket
\be\label{bracket}
\{f,g\}_{(s)}:=I^{\alpha_{I}\beta_{I}}\partial_{\alpha_I}f\wedge\partial_{\beta_{I}}g\,,\qquad |I|=s-1\,.
\ee
In particular, note that the amplitude vanishes on \emph{any} conformally flat background if $s_{2}+s_{3}-s_{1}$ is odd.

The formula \eqref{umhvb3} holds for any choice of twistor representatives restricted to the unitary subsector. All the twistor integrals can be done to obtain an amplitude in momentum space by evaluating $\widetilde{\cM}_{3,-1}$ on the momentum eigenstates \eqref{momeig}. For $\tilde{h}_{1}$ and $h_{2},h_{3}$ of this form, the bracket acts as:
\begin{equation*}
\left\{h_{2},h_{3}\right\}_{(s_{23|1}+1)}=t_{2}^{s_{23|1}}\,t_{3}^{s_{23|1}}\left([2\,3]+\Lambda\,\left\la\frac{\partial}{\partial k_2}\frac{\partial}{\partial k_3}\right\ra\right)^{s_{23|1}}\,h_{2}\,h_{3}\,,
\end{equation*}
and all integrals can be performed analogously to the full CHS case to give:
\begin{multline}\label{umhvb4}
\widetilde{\cM}_{3,-1}(-s_1,+s_{2},+s_{3})=\Lambda^{s_{1}-1}\,\widetilde{\cN}^{(s_1,s_2,s_3)}\,[2\,3]^{2s_{1}+1}\,[1\,2]^{s_{2}-s_{1}-s_{3}}\,[3\,1]^{s_{3}-s_{1}-s_{2}} \\
\left([2\,3]+\Lambda\,\left\la\frac{\partial}{\partial k_2}\frac{\partial}{\partial k_3}\right\ra\right)^{s_{23|1}}\,\delta^{4}\!\left(\sum_{i=1}^{3}|i\ra [i|\right)\,.
\end{multline}
The differential operator appearing in the second line can be simplified by noting that when acting on the four-momentum conserving delta function, $\la\frac{\partial}{\partial k_{2}}\frac{\partial}{\partial k_3}\ra$ is equivalent to $[23] \Box_{P}$, where $\Box_P$ is the d'Alembertian with respect to the total 4-momentum:
\begin{equation*}
 \Box_{P}:=\frac{\partial}{\partial P_{AA'}}\frac{\partial}{\partial P^{AA'}}\,, \qquad P^{AA'}:=\sum_{i=1}^{n}k_{i}^{A}\tilde{k}_{i}^{A'}\,.
\end{equation*}
This reduces the $\overline{\mbox{MHV}}$ answer for the unitary subsector to:
\be\label{umhvb5}
 \widetilde{\cM}_{3,-1}(-s_1,+s_{2},+s_{3})=\Lambda^{s_{1}-1}\,\frac{\widetilde{\cN}^{(s_1,s_2,s_3)}\,[2\,3]^{s_{1}+s_{2}+s_{3}}}{[1\,2]^{s_{1}+s_{3}-s_{2}}\,[3\,1]^{s_{1}+s_{2}-s_{3}}}\left(1+\Lambda\,\Box_{P}\right)^{s_{23|1}}\,\delta^{4}\!\left(P\right)\,.
\ee
The presence of the differential operator acting on the momentum-conserving delta function can be viewed as a consequence of calculating the `amplitude' on a conformally flat, rather than Minkowski, background~\cite{Adamo:2012nn}.

In~\cite{Joung:2015eny,Beccaria:2016syk}, the definition of S-matrix for CHS theory was taken to have all external states in the unitary subsector, evaluated on a Minkowski background. Having already restricted to the unitary subsector in \eqref{umhvb5}, the flat space limit is simply given by $\Lambda\rightarrow 0$:
\be\label{fslim1}
\lim_{\Lambda\rightarrow 0}\widetilde{\cM}_{3,-1}(-s_1,+s_{2},+s_{3})=0\,,
\ee
with the amplitude vanishing at $O(\Lambda^{s_{1}-1})$. So \emph{all} three-point $\overline{\mbox{MHV}}$ amplitudes of CHS theory vanish when restricted to the unitary subsector on Minkowski space.

\medskip

The generating functional \eqref{MHV1} for MHV amplitudes on a conformal gravity background is easily restricted to the unitary subsector:
\begin{multline}\label{uMHVgf}
\widetilde{I}^{\mathrm{MHV}}=\sum_{|I|=0}^{\infty}\int\frac{\d^{8}X}{\mathrm{vol}\;\GL(2,\C)}(1\,2)^{2}\,\D\sigma_{1}\,\D\sigma_{2}\,\left\la \cZ(x,\sigma_{1}),\cZ(x,\sigma_{2})\right\ra^{2|I|} \\
\tilde{h}^{(|I|+1)}_{1}(\cZ(x,\sigma_1))\,\tilde{h}^{(|I|+1)}_{2}(\cZ(x,\sigma_2))\,,
\end{multline}
where $\la \cZ(x,\sigma_{1}),\cZ(x,\sigma_{2})\ra = I_{\alpha\beta} \cZ^{\alpha}(x,\sigma_1) \cZ^{\beta}(x,\sigma_2)$. After choosing a definite spin for the negative helicity external states the perturbative expansion is operationalized using:
\be\label{upert}
\cZ^{\alpha}(x,\sigma)\rightarrow  \frac{1}{2\pi\im}\int_{\CP^1}\frac{\D\sigma '}{(\sigma\sigma')}\frac{(\xi\sigma)^2}{(\xi\sigma')^{2}}\,I^{\beta\alpha}\,\frac{\partial h^{(2)}(Z(x,\sigma'))}{\partial Z^{\beta}(x,\sigma')}\,.
\ee
Applying the matrix tree theorem with dual twistor wave functions results in
\begin{multline}\label{umhv1}
 \widetilde{\cM}_{n,0}=\int\frac{\d^{8}X}{\mathrm{vol}\;\GL(2,\C)}(1\,2)^{2}\,\D\sigma_{1}\,\D\sigma_{2}\bigg[\left\la Z(x,\sigma_1),Z(x,\sigma_2)\right\ra^{2s-2}\,\left|\HH^{12}_{12}\right| \\
+(2s-2)\Lambda\sum_{i=3}^{n} \left|\HH^{12i}_{12i}\right| \left\la Z(x,\sigma_1),Z(x,\sigma_2)\right\ra^{2s-3}\, t_{i}W_{i}\cdot Z(x,\sigma_2)\, \m^{1}_{i} + (1\leftrightarrow 2) \\
+\cdots+ \Lambda^{2s-2}(2s-2)!\sum_{\substack{i_{1},\ldots,i_{s-1} \\ j_{1},\ldots,j_{s-1}}}\left|\HH^{12i_{1}\cdots j_{s-1}}_{12i_{1}\cdots j_{s-1}}\right|\prod_{a=1}^{s-1}\m^{1}_{i_a}\m^{2}_{j_a}\,t_{i_a}t_{j_a}\left[W_{i_a},W_{j_a}\right]\bigg]\\
\times \e^{\im\cP\cdot X}\,(t_{1}t_{2})^{s+4}\prod_{k=1}^{n}\frac{\d t_{k}}{t_{k}^{2}}\,,
\end{multline}
where $\HH$ is the $n\times n$ matrix
\be\label{Hmat}
\HH_{ij}= (-1)^{|i-j|} t_{i}t_{j}\,\left[W_{i},W_{j}\right]\,\frac{\sqrt{\D\sigma_{i}\D\sigma_{j}}}{(i\,j)}\,, \quad \mbox{ for } i\neq j\,,
\ee 
\begin{equation*}
\HH_{ii}=-t_{i}\,\D\sigma_{i}\,\sum_{j\neq i} (-1)^{|i-j|} t_{j}\frac{[W_{i}, W_{j}]}{(i\,j)}\frac{(\xi\,j)^2}{(\xi\,i)^2}\,,
\end{equation*}
and $\m^{i}_{j}$ is given by \eqref{mcont}.

As it stands, little can be said about the flat space limit of \eqref{umhv1} because the moduli integrals over $X^{\alpha A}$ remain to be performed, and there are no explicit overall powers of $\Lambda$. To proceed, we follow the method of~\cite{Adamo:2013tja} to partially evaluate the moduli integrals. First, the scaling parameters $t_{i}$ can be absorbed into the homogeneous coordinates $\sigma_i$, at the expense of all $\CP^1$ integrals becoming $\C^{2}$ integrals: $\d t_{i} \D\sigma_{i} \rightarrow \d^{2}\sigma_i$. We can also use the fact that $Z^{\alpha}(x,\sigma)=X^{\alpha A}\sigma_{A}$ to write
\begin{equation*}
 \left\la Z(x,\sigma_1), Z(x,\sigma_2)\right\ra=(1\,2)\, I_{\alpha\beta}\,X^{\alpha}_{A}\,X^{\beta A}:= (1\,2)\,X^2\,.
\end{equation*}
This allows us to rewrite the MHV amplitude as
\begin{multline}\label{umhv2}
 \widetilde{\cM}_{n,0}=\int\frac{\d^{8}X}{\mathrm{vol}\;\GL(2,\C)}(1\,2)^{2}\,\d^{2}\sigma_{1}\,\d^{2}\sigma_{2}\bigg[(1\,2)^{2s-2} (X^2)^{2s-2}\,\left|\HH^{12}_{12}\right| \\
+(2s-2)\Lambda\sum_{i=3}^{n} \left|\HH^{12i}_{12i}\right|\, (1\,2)^{2s-3}\,(X^2)^{2s-3}\, (W_{i}\!\cdot\!X\cdot\sigma_2) \, \m^{1}_{i} + (1\leftrightarrow 2) \\
+\cdots+ \Lambda^{2s-2}(2s-2)!\sum_{\substack{i_{1},\ldots,i_{s-1} \\ j_{1},\ldots,j_{s-1}}}\left|\HH^{12i_{1}\cdots j_{s-1}}_{12i_{1}\cdots j_{s-1}}\right|\prod_{a=1}^{s-1}\m^{1}_{i_a}\m^{2}_{j_a}\,\left[W_{i_a},W_{j_a}\right]\bigg]\,\e^{\im\cP\cdot X}\,,
\end{multline}
with generalized momentum $\cP=\sum_{i}W_{i}\sigma_i$.

It is clear that the only potentially non-vanishing contribution in the flat space limit is the first term in \eqref{umhv2}, since every other term is of $O(\Lambda)$:
\be\label{mhvfs}
\lim_{\Lambda\rightarrow 0}\widetilde{\cM}_{n,0}=\lim_{\Lambda\rightarrow 0}\int\frac{\d^{8}X}{\mathrm{vol}\;\GL(2,\C)}(1\,2)^{2s}\,\left|\HH^{12}_{12}\right|\,\d^{2}\sigma_{1}\,\d^{2}\sigma_{2}\,(X^2)^{2s-2}\,\e^{\im\cP\cdot X}\,.
\ee
Note that $X^2$ can be seen as a differential operator acting on $\e^{\im\cP\cdot X}$ of the form
\begin{equation*}
 X^{2}\leftrightarrow \Box:=\frac{1}{(1\,2)}\left\la\frac{\partial}{\partial W_{1}},\,\frac{\partial}{\partial W_{2}}\right\ra\,, 
\end{equation*}
and \eqref{mhvfs} can then be written as
\be\label{mhvfs2}
\lim_{\Lambda\rightarrow 0}\int\frac{\d^{8}X}{\mathrm{vol}\;\GL(2,\C)}(1\,2)^{2s}\,\left|\HH^{12}_{12}\right|\,\d^{2}\sigma_{1}\,\d^{2}\sigma_{2}\,\Box^{2s-2}\,\e^{\im\cP\cdot X}\,.
\ee
The moduli integrals can now be performed to leave
\be\label{mhvfs3}
\lim_{\Lambda\rightarrow 0}\int\frac{\d^{2}\sigma_{1}\,\d^{2}\sigma_{2}}{\mathrm{vol}\;\GL(2,\C)}(1\,2)^{2s}\,\left|\HH^{12}_{12}\right|\,\Box^{2s-2}\,\delta^{8}\!\left(\cP\right)\,,
\ee
since the only $X$-dependence was in the exponential. Now, following~\cite{Adamo:2013tja}, integrate by parts twice to find
\begin{multline}\label{mhvfs4}
 \int\frac{\d^{2}\sigma_{1}\,\d^{2}\sigma_{2}}{\mathrm{vol}\;\GL(2,\C)}(1\,2)^{2s}\,\left|\HH^{12}_{12}\right|\,\Box^{2s-2}\,\delta^{8}\!\left(\cP\right) \\
 =\Lambda \int\frac{\d^{2}\sigma_{1}\,\d^{2}\sigma_{2}}{\mathrm{vol}\;\GL(2,\C)}\,(1\,2)^{2s}\,\bigg[\sum_{i}\frac{(1\,2)^2}{(1\,i)^{2}\,(2\,i)^{2}} \left|\HH^{12i}_{12i}\right| \\
 +\sum_{i,j}\frac{(\xi\,1)^{2}(i\,2)(j\,2)+(\xi\,2)^{2}(i\,1)(j\,1)}{(1\,i)(2\,i)(1\,j)(2\,j)(\xi\,i)(\xi\,j)}\left|\HH^{12i}_{12j}\right|\bigg]\,\Box^{2s-3}\delta^{8}\!\left(\cP\right)\,.
\end{multline}
In particular, an overall power of $\Lambda$ appears which was obscured in \eqref{mhvfs}-\eqref{mhvfs3}. 

Therefore, all MHV amplitudes on a conformal gravity background vanish in the flat space limit upon restricting to the unitary subsector:
\be\label{mhvfs5}
\lim_{\Lambda\rightarrow 0}\widetilde{\cM}_{n,0}=0\,,
\ee
with the zero appearing as $O(\Lambda)$. Although slightly less obvious than the $\overline{\mbox{MHV}}$ case, this provides an infinite set of amplitudes which confirm the conjecture of~\cite{Joung:2015eny,Beccaria:2016syk} that the S-matrix of CHS theory restricted to the unitary subsector is trivial.

\medskip

Given the complexity of the MHV amplitude formulae, it is useful to consider the $n=3$ case, which should be the parity conjugate of the $\overline{\mbox{MHV}}$ amplitude. For $n=3$ only a single iteration of the perturbative expansion \eqref{upert} is needed, which can act on the wavefunctions or the bare $Z^{\alpha}$ to give:
\begin{multline}\label{3mhv1}
\widetilde{\cM}_{3,0}=\int\frac{\d^{8}X}{\mathrm{vol}\;\GL(2,\C)}(1\,2)^{2}\,\D\sigma_{1}\,\D\sigma_{2}\left(\frac{\D\sigma_{3}\,(\xi\,1)^2}{(1\,3)\,(\xi\,3)^2}\,\left\la Z(\sigma_1),Z(\sigma_2)\right\ra^{2s-2}\,\left\{h_{3},\tilde{h}_{1}\right\}_{(2)}\,\tilde{h}_{2} \right.\\
\left.+(2s-2)\Lambda\,\frac{\D\sigma_{3}\,(\xi\,1)^2}{(1\,3)\,(\xi\,3)^2}\,\left\la Z(\sigma_1),Z(\sigma_2)\right\ra^{2s-3}\, Z(\sigma_2)\cdot\frac{\partial h_{3}}{\partial Z(\sigma_3)}\,\tilde{h}_{1}\tilde{h}_{2}\right)\,+\,(1\leftrightarrow 2)\,,
\end{multline}
where the choice of representative wavefunctions has been left arbitrary. Let us treat the first and second line (along with their $(1\leftrightarrow 2)$ partners) separately. The first line can be rewritten as:
\begin{equation*}
\int\frac{\d^{8}X}{\mathrm{vol}\;\GL(2,\C)}(1\,2)^{2s}\,(X^{2})^{2s-2}\D\sigma_{1}\,\D\sigma_{2}\frac{\D\sigma_{3}\,(\xi\,1)^2}{(1\,3)\,(\xi\,3)^2}\left[\frac{\partial h_{3}}{\partial Z(\sigma_3)},\frac{\partial\tilde{h}_{1}}{\partial Z(\sigma_1)}\right] \tilde{h}_{2} + (1\leftrightarrow 2)\,,
\end{equation*} 
and this can be considerably simplified by using the linearity of $Z(\sigma_i)=Z(x,\sigma_i)$. In particular, 
\begin{equation*}
\frac{\partial\tilde{h}_{1}}{\partial Z^{\alpha}(\sigma_1)}=\frac{\sigma_{2}^{B}}{(1\,2)}\frac{\partial\tilde{h}_{1}}{\partial X^{\alpha B}}\,,
\end{equation*}
and we can integrate by parts with respect to $X$. After some straightforward manipulations, the contributions from the first line of \eqref{3mhv1} can be recast as
\be\label{3mhv2}
2\Lambda (2s-2)\int\frac{\d^{8}X}{\mathrm{vol}\;\GL(2,\C)}(1\,2)^{2s}\,(X^{2})^{2s-3}\D\sigma_{1}\,\D\sigma_{2}\frac{\D\sigma_{3}\,(\xi\,1)^2\,\sigma_{2}^{B}}{(1\,3)\,(1\,2)\,(\xi\,3)^2}\frac{\partial h_{3}}{\partial\sigma^{B}_{3}}\tilde{h}_{1}\,\tilde{h}_{2} +(1\leftrightarrow 2)\,.
\ee
Now we can integrate by parts once more with respect to $\sigma_3$; after adding the $(1\leftrightarrow 2)$ contribution and applying the Schouten identity we are left with:
\be\label{3mhv3}
2\Lambda(2s-2)\int\frac{\d^{8}X}{\mathrm{vol}\;\GL(2,\C)}(X^2)^{2s-3}\frac{(1\,2)^{2s+2}}{(1\,3)^{2}\,(2\,3)^{2}}\,\D\sigma_{1}\,\D\sigma_{2}\,\D\sigma_{3}\,h_{3}\,\tilde{h}_{1}\,\tilde{h}_{2}\,.
\ee

As for the second line of \eqref{3mhv1}, a similar simplification arises upon writing the operator $Z(\sigma_2)\cdot\frac{\partial h_3}{\partial Z(\sigma_3)}$ as $\sigma_2\cdot\frac{\partial h_3}{\partial \sigma_3}$ and then integrating by parts with respect to $\sigma_3$. After adding the $(1\leftrightarrow 2)$ contribution and applying the Schouten identity one obtains \eqref{3mhv3} but without the factor of $2$ and with the opposite sign. Hence, the 3-point amplitude can be rewritten as:
\be\label{3mhv4}
\widetilde{\cM}_{3,0}=\Lambda(2s-2)\int\frac{\d^{8}X}{\mathrm{vol}\;\GL(2,\C)}(X^2)^{2s-3}\frac{(1\,2)^{2s+2}}{(1\,3)^{2}\,(2\,3)^{2}}\,\D\sigma_{1}\,\D\sigma_{2}\,\D\sigma_{3}\,h_{3}\,\tilde{h}_{1}\,\tilde{h}_{2}\,
\ee
for arbitrary wavefunctions. This simplification is special to the three-points; for $n>3$ the integrations by parts will create myriad new terms in the expansion of the amplitude. 

Upon inserting momentum eigenstates into \eqref{3mhv4} and performing all integrations, we obtain the momentum space expression
\be\label{3mhvf}
\widetilde{\cM}_{3,0}=(2s-2)\Lambda\,\frac{\la1\,2\ra^{2s+2}}{\la1\,3\ra^{2}\,\la2\,3\ra^{2}}\,\left(1+\Lambda\Box_{P}\right)^{2s+3}\,\delta^{4}(P)\,.
\ee 
This is the parity conjugate of \eqref{umhvb5} in the special case where two particles have the same helicity and the third is spin two, apart from the overall normalization.
For the appropriate choice of this normalisation the resulting action will thus be parity invariant however it is not generically so. 

%%%%%%%%%%%%%%%%%%%%%%%%%%%%%%%%%%%%%%%%%%
%%%%%%%%%%%%%%%%%%%%%%%%%%%%%%%%%%%%%%%%%%

\section{Discussion}
\label{Dis}

It is useful to discuss the unitary sector theory in a more general fashion and to compare with 
some of the known results from the literature. In the gravitational case, as discussed in~\cite{Maldacena:2011mk}, 
%by imposing the vanishing of the component of the derivative of the metric normal to a totally geodesic 
%surface,
 one can relate, at the level of tree-diagrams, ordinary gravity in the presence of a cosmological
constant to conformal gravity. In particular the dimensionless coupling of conformal gravity, $c_W$, is
related to the dimensionless quantity appearing in gravity, $c_E$, here for the AdS case, by
\be
c_W=\tfrac{1}{4}c_E=\frac{1}{8}\left(\frac{R^2}{8\pi G_2}\right)\,,
\ee
where $R\sim \Lambda^{-1/2}$ is the radius of the AdS space and $G_2$ is the usual gravitational constant. 

For the CHS theory we might hope to find analogous behaviour, which is to say that upon truncation to the unitary 
sector we find a massless higher spin theory on an AdS (or dS) background where the dimensionful 
couplings are related to the dimensionless CHS parameters through powers of $\Lambda$. 
This corresponds
to the powers of $\Lambda$ appearing in the computation of the cubic amplitudes \eqref{umhvb5} and we see 
that the interactions involving a spin-$s$ negative helicity excitation scale as $\Lambda^{s-1}$. 
Alternatively, 
given dimensionless couplings $c_{(s)}$ for the cubic interactions involving a negative helicity spin $s$ field in the CHS theory, we expect
\be
G_s\sim \frac{1}{c_{(s)}\, \Lambda^{s-1}}\,.
\ee
for $G_s$ the analogous dimensionful coupling in massless higher spin theory.
%It would be certainly 
%interesting to attempt to better understand this putative massless theory involving all spins $s>0$ both for
%the just the self-dual theory and for its parity invariant completion. 

Given this, it is unsurprising that the flat-space limit of the cubic couplings vanishes -- it simply follows from the 
definition of $G_s$ -- however non-vanishing answers can be obtained when the amplitudes are normalized by
certain powers of the cosmological constant \emph{before} the flat space limit is taken. This of course can be 
done in the case of conformal gravity and results in Einstein's theory expanded around flat space; in particular
it has been shown~\cite{Adamo:2012nn,Adamo:2012xe,Adamo:2013tja} to reproduce the MHV amplitude formula of Hodges~\cite{Hodges:2012ym}. 
%From \eqref{fsl4} it follows that implementing this rescaling at the level of the SD CHS action appears to define a massless higher-spin theory about flat space.
 For the unitary sector CHS $\overline{\mbox{MHV}}$ amplitudes, this normalisation is easily read off from \eqref{umhvb5}:
\be\label{fsl1}
\lim_{\Lambda\rightarrow 0} \frac{\widetilde{\cM}_{3,-1}(-s_1,+s_{2},+s_{3})}{\Lambda^{s_1 -1}}=\widetilde{\cN}^{(s_1,s_2,s_3)}\,\delta^{4}(P)\,\frac{[2\,3]^{s_{1}+s_{2}+s_{3}}}{[1\,2]^{s_{1}+s_{3}-s_{2}}\,[3\,1]^{s_{1}+s_{2}-s_{3}}}\,.
\ee
This result is the unique combination of spinor invariants compatible with the $\overline{\mbox{MHV}}$ helicity configuration, four-momentum
conservation (which emerges in the flat space limit), and Poincar\'e invariance~\cite{Benincasa:2007xk}. 

For the MHV amplitudes with two negative helicity spin $s$ particles and $n-2$ positive helicity spin two particles, the appropriate normalisation is to divide by a single power of $\Lambda$ for any $n$. This reflects the fact that such amplitudes represent the coupling of the CHS fields to a conformal gravity background only. It follows that after normalisation, the resulting flat space amplitude can be given as an expression purely in terms of on-shell four-momenta:
\be\label{fsl2}
\lim_{\Lambda\rightarrow0} \frac{\widetilde{\cM}_{n,0}}{\Lambda}\propto \delta^{4}(P)\,\frac{\la1\,2\ra^{2s+2}}{\la1\,i\ra^2 \, \la 2\,i\ra^{2}}\,\left|\Phi^{12i}_{12i}\right|\,,
\ee
where the $n\times n$ matrix $\Phi$ is the Hodges matrix~\cite{Hodges:2012ym}:
\begin{equation*}
 \Phi_{ij}= \frac{[i\,j]}{\la i\,j\ra}\,, ~ \mbox{ for } i\neq j\,,~~~{\rm and}~~~
\Phi_{ii}=-\sum_{j\neq i}\frac{[i\,j]}{\la i\,j\ra}\frac{\la\xi\,j\ra^2}{\la\xi\,i\ra^2}\,.
\end{equation*}
In reaching \eqref{fsl2} one must invoke various properties of the Hodges matrix along with the twistor `gauge' invariance of the amplitude itself.

\medskip

Upon restricting the self-dual action to the unitary subsector one notes that it splits into an infinite sum of `unitary' actions:
\be\label{fsl3}
S_{\mathrm{SD}}[h,\tilde{h}]=\sum_{s=1}^{\infty}\Lambda^{s-1}\,S^{(s)}_{\mathrm{SD}}[h,\tilde{h}^{(s)}]\,,
\ee
where
\begin{multline}\label{fsl4}
S^{(s)}_{\mathrm{SD}}[h,\tilde{h}^{(s)}]=\int_{\CPT}\D^{3}\cZ\,\wedge\,\tilde{h}^{(s)}\wedge\bigg(\dbar\,h^{(s)} \\
+\frac{(s-1)!}{(2s-2)!}\sum_{s_3=1}^{s}\sum_{s_2=1+s-s_3}^{\infty}\widetilde{\cN}^{(s,s_2,s_3)}\left\{h^{(s_2)},h^{(s_3)}\right\}_{(s_2+s_3-s-1)}\bigg)\,,
\end{multline}
with the bracket $\{\cdot,\cdot\}_{(k)}$ defined as in \eqref{bracket}.

So for each $s$, the action $S^{(s)}_{\mathrm{SD}}$ contains a single negative helicity state of spin $s$ and an infinite tower of higher spin positive helicity states. These theories are purely self-dual, in the sense that their only tree-level scattering amplitudes are three-point $\overline{\mbox{MHV}}$ amplitudes, and they possess two derivative massless equations of motion on space-time. Furthermore, $\Lambda$ can be set to zero in \eqref{fsl4} without any problems. While it is not guaranteed that this is a consistent truncation for the HS case, \eqref{fsl4} appears to define an infinite family of non-conformal, chiral, flat space higher spin theories.

There are powerful no-go theorems, which are not applicable in the presence of a cosmological constant, that argue against the existence  of such flat space HS theories, notably the Weinberg low-energy theorem~\cite{Weinberg:1964ew} and the Coleman-Mandula theorem~\cite{Coleman:1967ad}, and so the existence of a $n$-point amplitude such as \eqref{fsl2} is puzzling. However, recently there has been renewed interest in the possibility of evading these theorems, in particular by the use of light-cone formulations, see~\cite{Bengtsson:1983pd,Bengtsson:1983pg,Bengtsson:1986kh} and~\cite{Fradkin:1991iy,Metsaev:1993ap,Metsaev:2005ar,Metsaev:2007rn,Bengtsson:2012jm,Bengtsson:2016alt,Bengtsson:2016jfk}, where it has been shown that there exists
a non-trivial chiral theory with cubic interactions which, while it is non-parity invariant and non-unitary, is nonetheless consistent~\cite{Metsaev:1991mt,Metsaev:1991nb,Ponomarev:2016lrm}
\footnote{We are particularly indebted to Dmitry Ponomarev and Evgeny Skvortsov for discussions on this topic.} . 

For the twistor action of \eqref{fsl4}, defined by the flat limit of the SD sector, one can read off the properties of cubic couplings, $C^{\lambda_1, \lambda_2, \lambda_3}$, involving
helicities $\lambda_1, \lambda_2, \lambda_3$ that would appear in a corresponding light-cone Hamiltonian by examining the three-point amplitudes
\footnote{For a discussion of the match between the off-shell \textit{Lagrangian} couplings and three-point amplitudes see \cite{Conde:2016izb, Sleight:2016xqq}.}. The spinor-helicity formalism is in fact very closely related 
to the light-cone formalism and a direct translation is possible (\textit{e.g.}, \cite{Ananth:2012un, Ponomarev:2016cwi}). The $\overline{\mbox{MHV}}$ amplitude would correspond to an interaction with coupling 
$C^{-s_1, s_2, s_3}$ where there are $s_2+s_3-s_1$ powers of momentum in the light cone vertex and we have $s_1\geq s_2, s_3$ and $s_1\leq s_2+s_3$. Amongst these interactions are those involving helicities $s$, $-s$ and $s'<s$ and which would correspond to $s'$ derivatives in a covariant approach, however it is by no means clear that the twistor actions \eqref{fsl4} are equivalent to manifestly covariant space-time actions. Such interactions are also present in the light-cone formalism; indeed all helicity configurations are allowed in this formalism. 
 
In contradistinction, one immediate consequence of the structure of the twistor action is that it will not give rise to cubic interactions involving only positive helicities. Such interactions in Yang-Mills
theory would correspond to adding $F^3$ terms in the space-time action; such vertices would require a dimensionful coupling parameter and so violate conformal invariance. For gravity this would correspond to adding an $R^3$ term to the action, but such terms cannot arise from the quadratic-in-curvature action of Weyl gravity. Similarly, in our CHS theory such `Abelian' interactions are absent. These interactions are, however, present in the chiral higher-spin theory discussed in~\cite{Metsaev:1991mt, Metsaev:1991nb, Ponomarev:2016lrm}. Specifically, by considering the kinematic consistency of the quartic couplings one finds constraints on the cubic couplings, a solution to which is given by the very simple formula
\be
C^{\lambda_1, \lambda_2, \lambda_3}=\frac{(1-(-1)^{\kappa})\ell_P^{\kappa}}{2 \kappa!}
\ee
where $\kappa=\lambda_1+\lambda_2+\lambda_3-1$. 

In addition to the presence of the all positive helicity coupling, the explicit expression for $\lambda_1=-s_1$, $\lambda_2=s_2$, $\lambda_3=s_3$
differs from the normalisation $\widetilde{\cN}^{(s_1,s_2,s_3)}$ in \eqref{unorm}. The appearance of $\Gamma(s_2+s_3-s_1)$ is in common
but the remaining factorial terms are different and cannot be removed by rescaling the fields as they involve different spins. It is not clear that one
can satisfy the light-cone kinematic constraints with only couplings of the
form $C^{-s_1, s_2, s_3}$ if spin greater than two is allowed, which would seem to rule out
a potential match. However, it may be that the flat space limit requires better understanding, or the twistor description of SD CHS theory can be modified to include additional interactions.

%Such a discussion goes beyond the scope of this work, but one can note
%that the simplest missing interactions, those of the form $C^{0,0,s}$, which are essentially the coupling
%of currents to higher spin fields can be easily found. Such interactions could be added to the twistor
%CHS action by introducing a coupling between the self-dual background and a
%field of homogeneity degree $-2$, $\phi\in \Omega^{0,1}(\CPT, {\cal O}(-2))$:
%\be
%S_{\rm Scalar}=\int_{\CPT} \Omega\wedge \phi\wedge \bar\partial_f \phi~.
%\ee
%The linearised action implies that $\phi$ defines a space-time scalar $\Phi$ satisfying the usual
%conformal equation of motion 
%\be
%(\Box+\tfrac{1}{6}R) \Phi=0
%\ee
%while the non-linear terms will give rise to couplings between the positive helicity spin $s$ 
%fields and the scalar. 

While the self-dual theory appears to be self-consistent, we naturally wish to define the full parity invariant theory. 
As we have described, this is done by adding anti-self-dual interaction terms and we have shown that, 
at least for the case of positive helicity spin-two fields, the resulting cubic interactions give MHV amplitudes
which are the appropriate parity conjugates of the $\overline{\mbox{MHV}}$ amplitudes. As we saw in \eqref{fsl2}, all such MHV amplitudes have the same scaling in the flat space limit, so normalizing by a power of $\Lambda^{-1}$ yields non-vanishing flat space expressions. Since these are candidates for $n$-point amplitudes, one must reckon with more recent on-shell arguments~\cite{Benincasa:2007xk, Benincasa:2011pg, Fotopoulos:2010ay, Dempster:2012vw} as well as the traditional no-go theorems. The light-cone theory potentially eludes the grasp of these on-shell arguments as it allows four-point amplitudes which are not BCFW constructible \cite{Bengtsson:2016hss}.
It would be interesting to perform the same analysis for the higher-point MHV amplitudes we found in \eqref{fsl2}.

\acknowledgments

We thank S. Nakach, D.Ponomarev, E. Skvortsov, M.Taronna, M.Tsulaia, A.Tseytlin, and M. Vasiliev for many useful conversations and comments. We also thank A. Tseytlin, D. Ponomarev, and E. Skvortsov for providing comments on a draft. The work of TA is supported by an Imperial College Junior Research Fellowship. The work of PH and TMcL is supported in part by Marie Curie Grant CIG-333851.

\bibliography{CHS}
\bibliographystyle{JHEP}

\end{document}